# Implementation of the Spherical Coordinate Representation of Protein 3D Structures and its Applications Using FORTRAN 77/90 Language

**Vicente M. Reyes, Ph.D.***

**E-mail:** vmrsbi.RIT.biology@gmail.com

*work done at:

Dept. of Pharmacology, School of Medicine,
University of California, San Diego
9500 Gilman Drive, La Jolla, CA 92093-0636

&

Dept. of Biological Sciences, School of Life Sciences
Rochester Institute of Technology
One Lomb Memorial Drive, Rochester, NY 14623

**Abbreviations:** PDB, Protein Data Bank, OL, outer layer; IC, inner core; AAR, all-atom representation; DCRR, double-centroid reduced representation; BS, binding site; LBS, ligand binding site; FD, frequency distribution; FDMR, frequency distribution of maximum rho's

## 1. ABSTRACT:

In an earlier paper we described the representation of protein 3D structures in spherical coordinates (rho, phi, theta) and two of its potential applications, namely, the separation of the outer layer (OL) from the inner core (IC) of proteins, and analysis of protein surface topography as to protrusions and invaginations (Reyes, V.M., 2011; see also Reyes, V.M., 2009). In the present paper, we present several results demonstrating the performance success of the several FORTRAN 77 and 90 program source codes used in the implementation of the two aforementioned applications, as well as how to exactly implement both applications using the set of programs. In particular, we show here data that demonstrate the success of our procedure for the separation of the OL from the IC of proteins using a subset of the dataset from Laskowski et al. (1996). Using a theoretical model protein in the form of a scalene ellipsoid grid of points with and without an artificially constructed protrusion or invagination, we also show results demonstrating that protrusions and invaginations on the surface of proteins maybe predicted using our procedure. Referring to Figures 1 and 2 of Reyes, V.M. (2011), the nine propgram codes we present here and their respective functions are as follows: (1.) find_molec_centr.f: finds the x-, y- and z-coordinates of the proteins molecular (geometric) centroid; (2.) cart2sphere_degrees.f90: converts the Cartesian PDB coordinates of the protein to spherical coordinates, with phi ($\phi$) and theta ($\theta$) in degrees; (3.) cart2sphere_radians.f90: does the same thing as the second program, but with $\phi$ and $\theta$ in radians; (4.) spher2cart_degrees.f90: converts the protein atomic coordinates from spherical back to Cartesian, where the input $\phi$ and $\theta$ values are in degrees; (5.) spher2cart_radians.f90: does the same thing as the fourth program, but with $\phi$ and $\theta$ in radians; (6.) find_rho_cutoff.f: determines the rho ($\rho$) cut-off value for delineating the boundary between the OL and the IC of the protein; (7.) phi6_theta8_binning.f90: performs the binning of $\phi$ in six-degree increments and $\theta$ in eight-degree increments; (8.) phi10_theta10_binning.f90: performs the binning of $\phi$ and $\theta$ both in ten-degree increments; and (9.) bin_rho.f90: performs the binning of $\rho$ values for plotting the frequency distribution of maximum $\rho$ values (FDMR).

## 2. INTRODUCTION:

Most protein 3D structures are solved by x-ray crystallography (XRC), while a small subset are solved by nuclear magnetic resonance (NMR) spectroscopy; and all of them are deposited in the Protein Data Bank (PDB) , the world's main repository of protein 3D structures. In each case, the final product is a PDB file containing the (x,y,z) Cartesian coordinates of all the atoms in the protein (in the case of XRC, only non-hydrogen atoms in general are assigned coordinates). While it is known that close to 80% of all proteins are globular or roughly globular in shape – and thus are often called 'spheroproteins' – prior to our report in 2011, protein 3D structures had never been represented using spherical coordinates in order to take advantage of their spherical symmetry. In this 2011 paper (Reyes, V.M., 2011), we reported on our efforts at representing spheroprotiens using spherical ($\rho$, $\varphi$, $\theta$) coordinates, with the protein centroid as origin, and presented two of its potential applications. Here, we present the Fortran program source codes we used in our procedure, starting with the determination of the protein molecular (geometric) centroid and the transformation of its Cartesian coordinates to spherical coordinates. We also present here the Fortran source codes for the implementation of the two applications we presented in the aforementioned paper, namely, the separation of the protein outer layer (OL) from its iner core (IC), and the identification of protrusions and invaginations on the protein surface.

The source program codes presented in this work were written in either Fortran 77 (Holoien, M.O. & Behforooz, A., 1991; Mayo, W. & Cwiakala, M., 1994; and Nyhoff, L. & Leestma, S., 1996) or Fortran 90 ( Nyhoff, L. & Leestma, S., 1996 & 1999; Metcalf, M. & Reid, J.K., 1999; and Chapman, S.J., 1997). All computations were carried out on a UNIX computing environment with a Fortran compiler software. In order to apply the procedure in high-throughput batch mode, UNIX C-shell (Powers, S. et al., 2002; Anderson, G. & Anderson, P., 1986; and Birns, P. et al., 1985) as well as Perl (Tisdall, J., 2001; and Berman, J.J., 2007) scripts were written. In some complex cases, the scripts were constructed using text manipulation programming languages, sed & awk (Dougherty, D. & Robbins, A., 1997; and Aho et al., 1988).

## 3. DATASETS AND METHODS:

All protein 3D structures are encoded as plain text files consisting of roughly four parts, namely, a meta data section about the file itself and the protein; the main part, which contains the x-, y-, and z- coordinates of all the atoms in the protein arranges in specific columns; water molecules that were detected in close association with the protein; and a geometry file for any ligands or other molecules bound by the protein in question, as well as for some unusual amino acid side chain rotamers, if any had been detected. Protein 3D structures may be downloaded from the Protein Data Bank (PDB; Berman et al., 2000) - the main international repository for protein 3D structures - which may then be visualized on the computer screen using several available computer graphics software. To demonstrate the utility and effectivity of our procedure, the programs were originally applied to the dataset in Laskowski (1996), which is composed of 67 globular monomeric enzymes (spheroproteins) all complexed with their cognate ligand(s). The procedure was also applied to a “theoretical” protein made by constructing a 3D grid of points representing protein atoms in the shape of a acalene ellipsoid with center at the origin with the three major axes along the x-, y- and z-coordinate axes (Reyes, V.M., 2011). We recommend that the present paper be read in conjunction with Reyes, V.M. (2011) in order for the reader to see precisely how the programs presented here are implemented.

The minimum requirements in running these program source codes is a UNIX computing environment and a Fortran 77/90 compiler software as all Fortran program source codes must first be compiled before they are run. Application of the procedure to a large dataset of protein structures will be require a knowledge of scripting methods such as UNIX C-shell scripting and Perl programming, as well as that of sed and awk for text manipulation.

## 4. RESULTS AND DISCUSSION:

The Fortran program source codes presented in this paper are shown in the table below; page numbers on tre third column refer to the page in the present paper where the program starts. Program names ending in suffix “.f” are Fortran 77 codes, while those ending in “.f90” are Fortran 90 codes. Please refer to our previous publication, Reyes, V.M. (2011) for the implementation of these programs on actual datasets. Specifically, please refer to Figures 1 and 2 of said paper for a flowchart and illustration of how these programs are implemented.

**Table of Programs**

---

---

Program 1, cart2sphere_degrees.f90, transforms the Cartesian coordinates in the protein PDB file to spherical coordinates, ($\rho$, $\varphi$, $\theta$) with $\varphi$ and $\theta$ measured in degrees. Program 2, cart2sphere_radians.f90, does the same

thing as Program 1, but expresses φ and θ in radians. Program 3, spher2cart_degrees.f90, back-transforms the protein atom coordinates from spherical coordinates to Cartesian, where φ and θ in the sherical coordinates are in degrees; program 4, spher2cart_radians.f90, does the same, but where φ and θ in the sherical coordinates are in radians. Program 5, find_rho_cutoff.f, calculates the cut-off ρ value to be used in delineating the protein OL from the protein IC. Program 6, phi6_theta8_binning.f90, bins (partitions) φ in increments of 6 degrees and θ in increments of 8 degrees; program 7, phi10_theta10_binning.f90, does the same but in increments of 10 degrees for both φ and θ. We thus call binning by phi6_theta8_binning.f90 'fine binning,' and binning by phi10_theta10_binning.f90 'coarse binning.' Finally, program 8, bin_rho.f90, bins ρ in preparation for plotting the frequency distribution of maximum ρ's.

Transforming a protein PDB structure file in Cartesian coordinates to spherical coordinates involves use of the program cart2sphere_degrees.f90 or cart2sphere_radians.f90, depending on whether the user wants the angles φ and θ in the resulting spherically transformed file to be in degrees or radians. Table 1 shows such a transformation. The standard, original PDB file is shown on the bottom panel, while the spherically ransformed file is shown on the upper panel. Both files have identical first to fifth columns; the difference is that in the original PDB file, columns six, seven and eight are the x- y- and z- coordinates of the atoms, while in the spherically transformed file, they are the spherical coordfinates ρ, φ and θ. The latter columns do not concern us very much here.

In all cases, we made sure to check that our spherical transformations were correct by ransforming tghem back to Cartesian coordinates (using program spher2cart_degrees.f90) and made sure that the output files we obtained were identical to the Cartesian PDB files we started with.

### 4.1 Implementation in the Separation of the Outer Layer from the Inner Core of Proteins.

A significant majority of proteins fold in three-dmensions such that the hyprophilic sites in its constituent amino acid residues are exposed on the surface and the hydrophobic sites are buried in the interior of the folded protein. In an aquaeus environment like the cytoplasm of the cell, folding in this way achieves the lowest energy conformational state for the protein (see Reyes, V. M., 2015x and references cited therein). However, there are exceptions. For example, membrane proteins are embedded in a largely hydrophobic environment consisting of membrane phospholipids, and they fold so that large patches of hydrophilic residues are exposed on the surface (ibid.).

We implemented our procedure for separating the outer layer from the inner core of a protein using the programs reported here. A measure of success for this procedure would be to identify and calculate the percentage of hydrophilic residues versus hydrophobic residues in the outer layer (OL) as well as in the inner core (IC) after the separation has been performed. We selected six proteins randomly from the dataset of Laskowski et al. (1996) and performed the separation procedure. The identities of the six proteins we term Proteins A-F is shown in Table 2.

Figure 1, Panels A-C, show the results of these separation procedures. When the proteins are left in all-atom represenration and fine binning is used, the results are as shown graphically in Panel A. with associated numerical values shown in Table 3, Panel A. Note that in all six cases, the OL is more highly enriched with hydrophilic residues than is the IC, and that the IC is more highly enriched in hydrophobic residues than is the OL. The degrees of enrichment vary, but the pattern is constant: OL more enriched in hydrophilic moieties and IC more enriched in hydrophobic ones. This pattern is preserved even if we performed coarse binning instead, as shown in Panel B, with associated numerical values in Table 3, Panel B. The patterned is likewise preserved if we transformed the proteins to double-centroid reduced prepresentation (DCRR) and used coarse binning, as shown in Panel C, with associated numerical values in Table 3, Panel C. Fine binning were not perfoemed with the proteins in DCRR as they appear to be incompatible (data not shown). Overall, the data show that there are no exceptional cases in our test set of six proteins, e.g., membrane proteins, which would have reversed the observed hydrophobicity/hydrophilicity pattern in the OL and IC, as expected from the identities of these proteins.

The ρ cut-off value for the OL-IC separation was varied between 85%-98% of the maximum ρ ($\rho_{max}$) and a value of 95% seemed to be optimal, as the difference between %hydrophobicity and %hydrophilicity between the OL and IC seems to reach maximum near this value (Reyes, V.M., 2011).

**4.2 Implementation in the Analysis of the Surface Topography of Proteins.**

Next we implemented our procedure for detecting invaginations (depressions, indentations, concavities) and protrusions (bulges, bumps, proturberances) on the surface of a protein. We created a theoretical "model" protein in the form of a scalene ellipsoid composed of a grid of points 1.5 units (Å) apart along the axial directions. We then created three variants of this model protein, namely, a variant with: (a.) an invagination; (b.) a protrusion; and (c.) with both an invagination and a protrusion. All four variants were then transformed to spherical coordinates, parftitioned into bins and the FDMR plots constructed as previously described (Reyes, V. M., 2011). Figure 2, Panel A shows the FDMR plot for the original ellipsoid model protein. It reveals a single "clean" peak centered around the $18^{th}$ ρ interval corresponding to a frequency of about 290. The ellipsoid on the right shows the approxinate 3D shape of the model protein whose FDMR is shown to the left. Panel B shows the FDMR plot for the model protein with an invagination. While there is a single peak of similar magnitude and peak location, we see a subpeak on the lagging side (to the left) of the main peak. We take this to be diagnostic of the invagination on the protein, as shown on the figure on the right. Panel C shows the FDMR plot for the model protein with a protrusion. Again the large main peak is present, but this time there is a small but conspicuous subpeak on the leading side (to the right) of the main peak. This peak is thus diagnostic of the protrusion on the protein as shown on the figure to the right. Panel D shows the FDMR plot for the variant with both an invagination and a protrusion. The single large peak seen in the three previous cases is still there, but now the two subpeaks on either side of the main peak seen in the the two previous cases are now present. This confirms that the subpeak on the left is indeed indicative of an invagination, and that the subpeak on the right is indicative of a protrusion.

Table 4 shows the results of the surface analysis procedure for the six test proteins from the Laskowski data set. Ligand binding sites and/or active sites in the form of invaginations or clefts on their protein surfaces have been detected using our procedure (i.e., spherical transformation and binning followed by FDMR plots). In protein A (Thioredoxin reductase), 44 residues were predicted to be in and/or around the LBS using coarse binning, which was reduced to 33 residues using fine binning. In protein B (Prostatic acid phosphatase), 53 such residues were detected using coarse binning, which was reduced to 36 residues with fine binning. In protein C (Xylanase), five such residues were detected by coarse binning, with fine binning reducing that number to two residues. In protein D (Human neutrophil elastase), coarse binning detected eight such residues, and further reduced to four residues by fine binning. In protein E (Porphobilinogen deaminase), 41 such residues were detected using coarse binning, but fine binning reduced that to .21 residues. And finally, in protein F (Bacteriochlorophyll-A protein), coarse binning detecte 41 such residues with fine binning reducing it to 28 residues. Preliminary analyses indicate that the residues detected above are indeed located at or near the depressions on the respective protein surface where the bound ligands in the corresponding structures are indeed docked.

**4.3 Conclusion**

In conclusion, we feel that the nine Fortran programs presented in this paper do perform their overall function quite well. In addition we conclude that the use of spherical coordinates to represent protein 3D structures provides a valid and effective way to separate the protein OL from the IC (which, in turn, might find other useful applications), as well as for the analysis of protein surface topography. Our objective for the foreseeable future is to express all protein 3D structures deposited in the PDB in spherical coordinates, and then apply the OL-IC separation procedure and surface topographical analysis to all of them.

**5. ACKNOWLEDGMENT:**

This work was supported by an Institutional Research and Academic Career Development Award to the author, NIGMS/NIH grant number GM 68524. The author wishes to acknowledge the San Diego Supercomputer Center, the UCSD Academic Computing Services, and the UCSD Biomedical Library, for the help and support of their staff and personnel. He also acknowledges the Division of Research Computing at RIT, and computing resources from the Dept. of Biological Sciences, College of Science, at RIT.

## 6. REFERENCES:

Anderson, G. and Anderson, P. (1986), The Unix C Shell Field Guide, Prentice Hall (Publ.)

Aho, A.V., Kernighan, B.W. & Weinberger, P.J. (1988) The AWK Programming Language, Pearson (Publ.)

Berman, J.J. (2007) Perl Programming for Medicine and Biology (Series in Biomedical Informatics), Jones & Bartlett (Publ.)

Berman, H.M., Westbrook, J., Feng, Z., Gilliland, G., Bhat, T.N., Weissig, H., Shindyalov, I.N., Bourne, P.E. (2000) The Protein Data Bank, Nucleic Acids Research, 28: 235-242. (URL: www.rcsb.org)

Birns, P., Brown, P., Muster, J.C.C. (1985 ) UNIX for People, Prentice Hall (Publ.)

Chapman, S.J. (1997) FORTRAN 90/95 for Scientists and Engineers, McGraw-Hill Science/Engineering/Math (Publ.)

Cheguri, S. and Reyes, V.M., **"**A Database/Webserver for Size-Independent Quantification of Ligand Binding Site Burial Depth in Receptor Proteins: Implications on Protein Dynamics**",** J. Biomol. Struct. & Dyn., Book of Abstracts, Albany 2011: The 17th Conversation, June 14-18, 2011, Vol. 28 (6) June 2011, p. 1013

Dougherty, D. and Robbins, A. (1997), Sed & Awk (2nd Ed.), O'Reilly Media (Publ.)

Holoien, M.O., Behforooz, A. (1991) Fortran 77 for Engineers and Scientists (2nd Ed.), Brooks/Cole Pub Co. (Publ.)

Laskowski, R.A., Luscombe, N.M., Swindells, M.B., Thornton, J.M. (1996). Protein clefts in recognition and function. Prot Sci 5, 2438–2452.

Mayo, W., Cwiakala, M. (1994) Schaum's Outline of Programming With Fortran 77 (Schaum's Outlines), McGraw-Hill Education (Publ.)

Metcalf, M. and Reid, J.K. (1999) Fortran 90/95 Explained (2nd Ed.), Oxford University Press (Publ.)

Nyhoff, L. and Leestma, S. (1996) FORTRAN 77 for Engineers and Scientists with an Introduction to FORTRAN 90 (4th Ed.), Pearson (Publ.)

Nyhoff, L. and Leestma, S. (1996) FORTRAN 90 for Engineers and Scientists, Pearson (Publ.)

Nyhoff, L. and Leestma, S. (1999) Introduction to FORTRAN 90, ESource Series (2nd Ed.), Prentice Hall (Publ.)

Powers, S., Peek, J., O'Reilly, T., Loukides, M. (2002), Unix Power Tools (3rd Ed.), O'Reilly Media (Publ.)

Reyes, V.M., (2009) **"**Representing Protein 3D Structures in Spherical Coordinates – Two Applications: 1. Detection of Invaginations, Protrusions, and Potential Ligand Binding Sites; and 2. Separation of Protein Hydrophilic Outer Layer from the Hydrophobic Core ", J. Biomol. Struct. & Dyn., Book of Abstracts, Albany 2009: The 16th Conversation, June 16-20 2009, Vol. 26 (6), pp. 874-5

Reyes, V.M. (2011) "Representation of Protein 3D Structures in Spherical (ρ,φ,θ) Coordinates and Two of Its Potential Applications", Interdiscip. Sci.: Comput. Life Sci., 2011, Vol. 3, No. 3, pp. 161-174.

Tisdall, J. (2001) Beginning Perl for Bioinformatics 1st Edition, O'Reilly Media (Publ.)

**7. FIGURES and LEGENDS:**

---

---

**8. TABLES and LEGENDS:**

---

---

**9. PROGRAMS:**

**Program 1:**

**######### Start of Pogram "find_molec_centr.f" #########**

```
program  find_molecular_centroid


! c c c c c c c c c c c c c c c c c c c c c c c c c c c c c c c c c c c c c c
!                                                                            c
!   Author:  Vicente M. Reyes, Ph.D.                                         c
!            Dept. of Pharmacol., Skaggs Sch. of Pharm. & Pharm. Sci.        c
!          & Dept. of Integrative Biosci., S.D. Supercomptr. Ctr.            c
!            La Jolla, CA  92093-0505  U.S.A.                                c
!                                                                            c
! c c c c c c c c c c c c c c c c c c c c c c c c c c c c c c c c c c c c c c

      character*30 left
      character*30 right
      integer count
      real x, y, z, sum_x, sum_y, sum_z, xc, yc, zc
```

```
      open (unit =1, file = "filei")
      open (unit =2, file = "fileo")

      count = 0
      sum_x = 0.00
      sum_y = 0.00
      sum_z = 0.00

888   read(1,100,end=333) left, x, y, z, right
100   format(A30, f8.3, f8.3, f8.3, A30)

      count = count + 1
      sum_x = sum_x + x
      sum_y = sum_y + y
      sum_z = sum_z + z

      go to 888

333   continue

      xc  = sum_x/count
      yc  = sum_y/count
      zc  = sum_z/count

      write (2,200) xc, yc, zc
200   format(f10.5, f10.5, f10.5)

!     print*, "x = ", xc
!     print*, "y = ", yc
!     print*, "z = ", zc

      close(2)
      close(1)

      stop
      end
```

**##########    End of Pogram  "find_molec_centr.f"  ##########**

**Program 2:**

**##########    Start of Pogram   "cart2sphere_degrees.f90"   ##########**

```
program C2S

! c c c c c c c c c c c c c c c c c c c c c c c c c c c c c c c c c c c c c
!                                                                         c
!   Author:  Vicente M. Reyes, Ph.D.                                      c
!            Dept. of Pharmacol., Skaggs Sch. of Pharm. & Pharm. Sci.     c
!         &  Dept. of Integrative Biosci., S.D. Supercomptr. Ctr.         c
!            La Jolla, CA  92093-0505  U.S.A.                             c
!                                                                         c
! c c c c c c c c c c c c c c c c c c c c c c c c c c c c c c c c c c c c c

      character*30 left, right
      character#31 tag
```

```
      real x, y, z, pi, rho, phi, theta

      open (unit =1, file = "filei")
      open (unit =2, file = "fileo")

      pi = 3.1415926536

      tag = "SPHER(RHO,PHI,THETA) in degrees"

!****************************************************************************
! The output of this program has rho, phi and theta in degrees, not radians!
!****************************************************************************

888   read(1,100,end=333) left, x, y, z, right

100   format(A30, f8.3, f8.3, f8.3, A30)

      rho = sqrt(x**2 + y**2 + z**2)

      phi = ((ACOS(z/rho))*(180.0/pi))

      S = sqrt(x**2 + y**2)

      if ((x.gt.0.00).and.(y.ge.0.00)) then

      theta = ((ASIN(y/S))*(180.0/pi))

      elseif ((x.le.0.00).and.(y.gt.0.00)) then

      theta = ((pi - ASIN(y/S))*(180.0/pi))

      elseif ((x.lt.0.00).and.(y.le.0.00)) then

      theta = ((pi - ASIN(y/S))*(180.0/pi))

      elseif ((x.ge.0.00).and.(y.lt.0.00)) then

      theta = ((2*pi + ASIN(y/S))*(180.0/pi))

      endif

!!!!!!!!!!!!!!!!!!!!!!!!!!!!!!!!!!!!!!!!!!!!!!!!!!!!!!!!!!!!!!!!
!          /  ASIN(y/S)              if x>0, y>=0 (qdt. I)
! theta = |   pi - ASIN(y/S)         if x=<0, y>0 (qdt. II)
!         |   pi - ASIN(y/S)         if x<0, y=<0 (qdt. III)
!          \  2*pi + ASIN(y/S)       if x>=0, y<0 (qdt. IV)
!!!!!!!!!!!!!!!!!!!!!!!!!!!!!!!!!!!!!!!!!!!!!!!!!!!!!!!!!!!!!!!!


      write (2,101) left, rho, phi, theta, tag

101   format(A30, f14.8, f14.8, f14.8, A31)

      go to 888

333   continue

      close(2)
      close(1)

      stop
      end
```

**##########   End of Pogram   "cart2sphere_degrees.f90"   ##########**

**Program 3:**

**##########   Start of Pogram   "cart2sphere_radians.f90"  ##########**

```
program cartesian2spherical

! c c c c c c c c c c c c c c c c c c c c c c c c c c c c c c c c c c c c c
!                                                                         c
!   Author:  Vicente M. Reyes, Ph.D.                                      c
!            Dept. of Pharmacol., Skaggs Sch. of Pharm. & Pharm. Sci.     c
!          & Dept. of Integrative Biosci., S.D. Supercomptr. Ctr.         c
!            La Jolla, CA  92093-0505  U.S.A.                             c
!                                                                         c
! c c c c c c c c c c c c c c c c c c c c c c c c c c c c c c c c c c c c c

      character*30 left, right
      character*31 tag
      real x, y, z, pi, rho, phi, theta

      open (unit =1, file = "filei")
      open (unit =2, file = "fileo")

      pi = 3.1415926536

      tag = "SPHER(RHO,PHI,THETA) in radians"

!***************************************************************************
! The output of this program has rho, phi and theta in radians, not degrees!
!***************************************************************************

888   read(1,100,end=333) left, x, y, z, right

100   format(A30, f8.3, f8.3, f8.3, A30)

      rho = sqrt(x**2 + y**2 + z**2)

      phi = ACOS(z/rho)

      S = sqrt(x**2 + y**2)

      if ((x.gt.0.00).and.(y.ge.0.00)) then

      theta = ASIN(y/S)

      elseif ((x.le.0.00).and.(y.gt.0.00)) then

      theta = pi - ASIN(y/S)

      elseif ((x.lt.0.00).and.(y.le.0.00)) then

      theta = pi - ASIN(y/S)

      elseif ((x.ge.0.00).and.(y.lt.0.00)) then

      theta = 2*pi + ASIN(y/S)
```

```
      endif

!!!!!!!!!!!!!!!!!!!!!!!!!!!!!!!!!!!!!!!!!!!!!!!!!!!!!!!!!!!!!!!!
!          /  ASIN(y/S)              if x>0, y>=0 (qdt. I)
! theta = |   pi - ASIN(y/S)         if x=<0, y>0 (qdt. II)
!         |   pi - ASIN(y/S)         if x<0, y=<0 (qdt. III)
!          \  2*pi + ASIN(y/S)       if x>=0, y<0 (qdt. IV)
!!!!!!!!!!!!!!!!!!!!!!!!!!!!!!!!!!!!!!!!!!!!!!!!!!!!!!!!!!!!!!!!


      write (2,101) left, rho, phi, theta, tag

101   format(A30, f8.4, f8.5, f8.5, A30)

      go to 888

333   continue

      close(2)
      close(1)

      stop
      end
```

########## End of Pogram "cart2sphere_radians.f90" ##########

**Program 4:**

########## Start of Pogram "spher2cart_degrees.f90" ##########

```
program spherical2cartesian

! c c c c c c c c c c c c c c c c c c c c c c c c c c c c c c c c c c c c c c c c c c c c c c c
!                                                                  c
!   Author:  Vicente M. Reyes, Ph.D.                               c
!            Dept. of Pharmacol., Skaggs Sch. of Pharm. & Pharm. Sci.   c
!         &  Dept. of Integrative Biosci., S.D. Supercomptr. Ctr.      c
!            La Jolla, CA  92093-0505  U.S.A.                      c
!                                                                  c
! c c c c c c c c c c c c c c c c c c c c c c c c c c c c c c c c c c c c c c c c c c c c c c c

      character*30 resinfo
      character*10 misc,binID
      real x,y,z, pi, rho,phi,theta

      open (unit =1, file = "filei")
      open (unit =2, file = "fileo")

      misc = "1.00100.00"

      pi = 3.1415926536

!*********************************************************************************
! The input to this program should have rho, phi and theta in degrees, not radians!
!*********************************************************************************

888   read(1,100,end=333) resinfo,rho,phi,theta,binID
```

```
100   format(A30,f14.8,f14.8,f14.8,20x,A10)

      x = rho*SIN(phi*(pi/180.0))*COS(theta*(pi/180.0))

      y = rho*SIN(phi*(pi/180.0))*SIN(theta*(pi/180.0))

      z = rho*COS(phi*(pi/180.0))

!********************************************
!   x = rho*sin(phi)*cos(theta)
!   y = rho*sin(phi)*sin(theta)
!   z = rho*cos(phi)
!********************************************

      write (2,101) resinfo,x,y,z,misc,binID

101   format(A30,f8.3,f8.3,f8.3,x,A10,x,A10)


      go to 888

333   continue

      close(2)
      close(1)

      stop
      end
```

**########## End of Pogram "spher2cart_degrees.f90" ##########**

**Program 5:**

**########## Start of Pogram "spher2cart_radians.f90" ##########**

```
program spherical2cartesian

! c c c c c c c c c c c c c c c c c c c c c c c c c c c c c c c c c c c c c
!                                                                         c
!   Author:  Vicente M. Reyes, Ph.D.                                      c
!            Dept. of Pharmacol., Skaggs Sch. of Pharm. & Pharm. Sci.     c
!         &  Dept. of Integrative Biosci., S.D. Supercomptr. Ctr.         c
!            La Jolla, CA  92093-0505  U.S.A.                             c
!                                                                         c
! c c c c c c c c c c c c c c c c c c c c c c c c c c c c c c c c c c c c c

      character*30 resinfo
      character*10 misc,binID
      real x,y,z, rho,phi,theta

      open (unit =1, file = "filei")
      open (unit =2, file = "fileo")

      misc = "1.00100.00"

!*********************************************************************************
! The input to this program should have rho, phi and theta in radians, not degrees!
```

```
!**********************************************************************************

888   read(1,100,end=333) resinfo,rho,phi,theta,binID

100   format(A30,f14.8,f14.8,f14.8,20x,A10)

      x = rho*SIN(phi)*COS(theta)

      y = rho*SIN(phi)*SIN(theta)

      z = rho*COS(phi)

!*******************************************
!   x = rho*sin(phi)*cos(theta)
!   y = rho*sin(phi)*sin(theta)
!   z = rho*cos(phi)
!*******************************************

      write (2,101) resinfo,x,y,z,misc,binID

101   format(A30,f8.3,f8.3,f8.3,x,A10,x,A10)


      go to 888

333   continue

      close(2)
      close(1)

      stop
      end
```

**########## End of Pogram "spher2cart_radians.f90" ##########**

**Program 6:**

**########## Start of Pogram "find_rho_cutoff.f" ##########**

```
c  program name: find_rho_cutoff.f

c c c c c c c c c c c c c c c c c c c c c c c c c c c c c c c c c c c c c
c                                                                       c
c   Author:  Vicente M. Reyes, Ph.D.                                    c
c            Dept. of Pharmacol., Skaggs Sch. of Pharm. & Pharm. Sci.   c
c         &  Dept. of Integrative Biosci., S.D. Supercomptr. Ctr.       c
c            La Jolla, CA  92093-0505  U.S.A.                           c
c                                                                       c
c c c c c c c c c c c c c c c c c c c c c c c c c c c c c c c c c c c c c

      character*30 resinfo, binIDs
      real rho, phi, theta, red_rho

      open (unit =1, file = "filei")   !<-- xx.prot.cortex (composed of max_rhos)
      open (unit =2, file = "fileo")   !<-- max_rho's reduxced to red_rho's, set by
researcher
```

```
888   read(1,100,end=333)resinfo,rho,phi,theta,binIDs

100   format(A30, f14.8, f14.8, f14.8, A30)

      red_rho = 0.98000*rho

      write (2,100) resinfo,red_rho,phi,theta,binIDs

      go to 888

333   continue

      close(2)
      close(1)

      stop
      end
```

**########## End of Pogram "find_rho_cutoff.f" ##########**

**Program 7:**

**########## Start of Pogram "phi6_theta8_binning.f90" ##########**

```
program phi6_theta8_binning

! c c c c c c c c c c c c c c c c c c c c c c c c c c c c c c c c c c c c c c c c c c c c c
!                                                                         c
!   Author:  Vicente M. Reyes, Ph.D.                                      c
!            Dept. of Pharmacol., Skaggs Sch. of Pharm. & Pharm. Sci.     c
!         &  Dept. of Integrative Biosci., S.D. Supercomptr. Ctr.        c
!            La Jolla, CA  92093-0505  U.S.A.                             c
!                                                                         c
! c c c c c c c c c c c c c c c c c c c c c c c c c c c c c c c c c c c c c c c c c c c c c

!  Bins are 6 degrees in phi and 8 degrees in theta.
!  Phi goes from 0 to 180 deg, so 180/6 = 30 levels in phi.
!  Theta goes from 0 to 360 deg, so 360/8 = 45 levels in theta.
!  Total no. of bins = 30x45 = 1350 bins.

      character*30 left
      character*30 right
      real rho, phi, the

      open (unit =1, file = "filei")
      open (unit =3, file = "fileo")

123   read(1,200, end=111) left, rho, phi, the, right
200   format(A30, f14.8, f14.8, f14.8, A30)


!!!! body of 'if' statements start here

if  ((the.ge.0.00000000).and.(the.lt.8.00000000)) then
    if  ((phi.ge.0.00000000).and.(phi.lt.6.00000000)) then
         write (3,200) left, rho, phi, the, "bin = 0001"
    elseif  ((phi.ge.6.00000000).and.(phi.lt.12.00000000)) then
```

```
        write (3,200) left, rho, phi, the, "bin = 0002"
    elseif  ((phi.ge.12.00000000).and.(phi.lt.18.00000000)) then
        write (3,200) left, rho, phi, the, "bin = 0003"
    elseif  ((phi.ge.18.00000000).and.(phi.lt.24.00000000)) then
        write (3,200) left, rho, phi, the, "bin = 0004"
    elseif  ((phi.ge.24.00000000).and.(phi.lt.30.00000000)) then
        write (3,200) left, rho, phi, the, "bin = 0005"
    elseif  ((phi.ge.30.00000000).and.(phi.lt.36.00000000)) then
        write (3,200) left, rho, phi, the, "bin = 0006"
    elseif  ((phi.ge.36.00000000).and.(phi.lt.42.00000000)) then
        write (3,200) left, rho, phi, the, "bin = 0007"
    elseif  ((phi.ge.42.00000000).and.(phi.lt.48.00000000)) then
        write (3,200) left, rho, phi, the, "bin = 0008"
    elseif  ((phi.ge.48.00000000).and.(phi.lt.54.00000000)) then
        write (3,200) left, rho, phi, the, "bin = 0009"
    elseif  ((phi.ge.54.00000000).and.(phi.lt.60.00000000)) then
        write (3,200) left, rho, phi, the, "bin = 0010"
    elseif  ((phi.ge.60.00000000).and.(phi.lt.66.00000000)) then
        write (3,200) left, rho, phi, the, "bin = 0011"
    elseif  ((phi.ge.66.00000000).and.(phi.lt.72.00000000)) then
        write (3,200) left, rho, phi, the, "bin = 0012"
    elseif  ((phi.ge.72.00000000).and.(phi.lt.78.00000000)) then
        write (3,200) left, rho, phi, the, "bin = 0013"
    elseif  ((phi.ge.78.00000000).and.(phi.lt.84.00000000)) then
        write (3,200) left, rho, phi, the, "bin = 0014"
    elseif  ((phi.ge.84.00000000).and.(phi.lt.90.00000000)) then
        write (3,200) left, rho, phi, the, "bin = 0015"
    elseif  ((phi.ge.90.00000000).and.(phi.lt.96.00000000)) then
        write (3,200) left, rho, phi, the, "bin = 0016"
    elseif  ((phi.ge.96.00000000).and.(phi.lt.102.00000000)) then
        write (3,200) left, rho, phi, the, "bin = 0017"
    elseif  ((phi.ge.102.00000000).and.(phi.lt.108.00000000)) then
        write (3,200) left, rho, phi, the, "bin = 0018"
    elseif  ((phi.ge.108.00000000).and.(phi.lt.114.00000000)) then
        write (3,200) left, rho, phi, the, "bin = 0019"
    elseif  ((phi.ge.114.00000000).and.(phi.lt.120.00000000)) then
        write (3,200) left, rho, phi, the, "bin = 0020"
    elseif  ((phi.ge.120.00000000).and.(phi.lt.126.00000000)) then
        write (3,200) left, rho, phi, the, "bin = 0021"
    elseif  ((phi.ge.126.00000000).and.(phi.lt.132.00000000)) then
        write (3,200) left, rho, phi, the, "bin = 0022"
    elseif  ((phi.ge.132.00000000).and.(phi.lt.138.00000000)) then
        write (3,200) left, rho, phi, the, "bin = 0023"
    elseif  ((phi.ge.138.00000000).and.(phi.lt.144.00000000)) then
        write (3,200) left, rho, phi, the, "bin = 0024"
    elseif  ((phi.ge.144.00000000).and.(phi.lt.150.00000000)) then
        write (3,200) left, rho, phi, the, "bin = 0025"
    elseif  ((phi.ge.150.00000000).and.(phi.lt.156.00000000)) then
        write (3,200) left, rho, phi, the, "bin = 0026"
    elseif  ((phi.ge.156.00000000).and.(phi.lt.162.00000000)) then
        write (3,200) left, rho, phi, the, "bin = 0027"
    elseif  ((phi.ge.162.00000000).and.(phi.lt.168.00000000)) then
        write (3,200) left, rho, phi, the, "bin = 0028"
    elseif  ((phi.ge.168.00000000).and.(phi.lt.174.00000000)) then
        write (3,200) left, rho, phi, the, "bin = 0029"
    elseif  ((phi.ge.174.00000000).and.(phi.lt.180.00000000)) then
        write (3,200) left, rho, phi, the, "bin = 0030"
    endif
elseif  ((the.ge.8.00000000).and.(the.lt.16.00000000)) then
    if  ((phi.ge.0.00000000).and.(phi.lt.6.00000000)) then
        write (3,200) left, rho, phi, the, "bin = 0031"
    elseif  ((phi.ge.6.00000000).and.(phi.lt.12.00000000)) then
```

```
            write (3,200) left, rho, phi, the, "bin = 0032"
        elseif  ((phi.ge.12.00000000).and.(phi.lt.18.00000000)) then
            write (3,200) left, rho, phi, the, "bin = 0033"
        elseif  ((phi.ge.18.00000000).and.(phi.lt.24.00000000)) then
            write (3,200) left, rho, phi, the, "bin = 0034"
        elseif  ((phi.ge.24.00000000).and.(phi.lt.30.00000000)) then
            write (3,200) left, rho, phi, the, "bin = 0035"
        elseif  ((phi.ge.30.00000000).and.(phi.lt.36.00000000)) then
            write (3,200) left, rho, phi, the, "bin = 0036"
        elseif  ((phi.ge.36.00000000).and.(phi.lt.42.00000000)) then
            write (3,200) left, rho, phi, the, "bin = 0037"
        elseif  ((phi.ge.42.00000000).and.(phi.lt.48.00000000)) then
            write (3,200) left, rho, phi, the, "bin = 0038"
        elseif  ((phi.ge.48.00000000).and.(phi.lt.54.00000000)) then
            write (3,200) left, rho, phi, the, "bin = 0039"
        elseif  ((phi.ge.54.00000000).and.(phi.lt.60.00000000)) then
            write (3,200) left, rho, phi, the, "bin = 0040"
        elseif  ((phi.ge.60.00000000).and.(phi.lt.66.00000000)) then
            write (3,200) left, rho, phi, the, "bin = 0041"
        elseif  ((phi.ge.66.00000000).and.(phi.lt.72.00000000)) then
            write (3,200) left, rho, phi, the, "bin = 0042"
        elseif  ((phi.ge.72.00000000).and.(phi.lt.78.00000000)) then
            write (3,200) left, rho, phi, the, "bin = 0043"
        elseif  ((phi.ge.78.00000000).and.(phi.lt.84.00000000)) then
            write (3,200) left, rho, phi, the, "bin = 0044"
        elseif  ((phi.ge.84.00000000).and.(phi.lt.90.00000000)) then
            write (3,200) left, rho, phi, the, "bin = 0045"
        elseif  ((phi.ge.90.00000000).and.(phi.lt.96.00000000)) then
            write (3,200) left, rho, phi, the, "bin = 0046"
        elseif  ((phi.ge.96.00000000).and.(phi.lt.102.00000000)) then
            write (3,200) left, rho, phi, the, "bin = 0047"
        elseif  ((phi.ge.102.00000000).and.(phi.lt.108.00000000)) then
            write (3,200) left, rho, phi, the, "bin = 0048"
        elseif  ((phi.ge.108.00000000).and.(phi.lt.114.00000000)) then
            write (3,200) left, rho, phi, the, "bin = 0049"
        elseif  ((phi.ge.114.00000000).and.(phi.lt.120.00000000)) then
            write (3,200) left, rho, phi, the, "bin = 0050"
        elseif  ((phi.ge.120.00000000).and.(phi.lt.126.00000000)) then
            write (3,200) left, rho, phi, the, "bin = 0051"
        elseif  ((phi.ge.126.00000000).and.(phi.lt.132.00000000)) then
            write (3,200) left, rho, phi, the, "bin = 0052"
        elseif  ((phi.ge.132.00000000).and.(phi.lt.138.00000000)) then
            write (3,200) left, rho, phi, the, "bin = 0053"
        elseif  ((phi.ge.138.00000000).and.(phi.lt.144.00000000)) then
            write (3,200) left, rho, phi, the, "bin = 0054"
        elseif  ((phi.ge.144.00000000).and.(phi.lt.150.00000000)) then
            write (3,200) left, rho, phi, the, "bin = 0055"
        elseif  ((phi.ge.150.00000000).and.(phi.lt.156.00000000)) then
            write (3,200) left, rho, phi, the, "bin = 0056"
        elseif  ((phi.ge.156.00000000).and.(phi.lt.162.00000000)) then
            write (3,200) left, rho, phi, the, "bin = 0057"
        elseif  ((phi.ge.162.00000000).and.(phi.lt.168.00000000)) then
            write (3,200) left, rho, phi, the, "bin = 0058"
        elseif  ((phi.ge.168.00000000).and.(phi.lt.174.00000000)) then
            write (3,200) left, rho, phi, the, "bin = 0059"
        elseif  ((phi.ge.174.00000000).and.(phi.lt.180.00000000)) then
            write (3,200) left, rho, phi, the, "bin = 0060"
        endif
    elseif  ((the.ge.16.00000000).and.(the.lt.24.00000000)) then
        if  ((phi.ge.0.00000000).and.(phi.lt.6.00000000)) then
            write (3,200) left, rho, phi, the, "bin = 0061"
        elseif  ((phi.ge.6.00000000).and.(phi.lt.12.00000000)) then
```

```
        write (3,200) left, rho, phi, the, "bin = 0062"
    elseif  ((phi.ge.12.00000000).and.(phi.lt.18.00000000)) then
        write (3,200) left, rho, phi, the, "bin = 0063"
    elseif  ((phi.ge.18.00000000).and.(phi.lt.24.00000000)) then
        write (3,200) left, rho, phi, the, "bin = 0064"
    elseif  ((phi.ge.24.00000000).and.(phi.lt.30.00000000)) then
        write (3,200) left, rho, phi, the, "bin = 0065"
    elseif  ((phi.ge.30.00000000).and.(phi.lt.36.00000000)) then
        write (3,200) left, rho, phi, the, "bin = 0066"
    elseif  ((phi.ge.36.00000000).and.(phi.lt.42.00000000)) then
        write (3,200) left, rho, phi, the, "bin = 0067"
    elseif  ((phi.ge.42.00000000).and.(phi.lt.48.00000000)) then
        write (3,200) left, rho, phi, the, "bin = 0068"
    elseif  ((phi.ge.48.00000000).and.(phi.lt.54.00000000)) then
        write (3,200) left, rho, phi, the, "bin = 0069"
    elseif  ((phi.ge.54.00000000).and.(phi.lt.60.00000000)) then
        write (3,200) left, rho, phi, the, "bin = 0070"
    elseif  ((phi.ge.60.00000000).and.(phi.lt.66.00000000)) then
        write (3,200) left, rho, phi, the, "bin = 0071"
    elseif  ((phi.ge.66.00000000).and.(phi.lt.72.00000000)) then
        write (3,200) left, rho, phi, the, "bin = 0072"
    elseif  ((phi.ge.72.00000000).and.(phi.lt.78.00000000)) then
        write (3,200) left, rho, phi, the, "bin = 0073"
    elseif  ((phi.ge.78.00000000).and.(phi.lt.84.00000000)) then
        write (3,200) left, rho, phi, the, "bin = 0074"
    elseif  ((phi.ge.84.00000000).and.(phi.lt.90.00000000)) then
        write (3,200) left, rho, phi, the, "bin = 0075"
    elseif  ((phi.ge.90.00000000).and.(phi.lt.96.00000000)) then
        write (3,200) left, rho, phi, the, "bin = 0076"
    elseif  ((phi.ge.96.00000000).and.(phi.lt.102.00000000)) then
        write (3,200) left, rho, phi, the, "bin = 0077"
    elseif  ((phi.ge.102.00000000).and.(phi.lt.108.00000000)) then
        write (3,200) left, rho, phi, the, "bin = 0078"
    elseif  ((phi.ge.108.00000000).and.(phi.lt.114.00000000)) then
        write (3,200) left, rho, phi, the, "bin = 0079"
    elseif  ((phi.ge.114.00000000).and.(phi.lt.120.00000000)) then
        write (3,200) left, rho, phi, the, "bin = 0080"
    elseif  ((phi.ge.120.00000000).and.(phi.lt.126.00000000)) then
        write (3,200) left, rho, phi, the, "bin = 0081"
    elseif  ((phi.ge.126.00000000).and.(phi.lt.132.00000000)) then
        write (3,200) left, rho, phi, the, "bin = 0082"
    elseif  ((phi.ge.132.00000000).and.(phi.lt.138.00000000)) then
        write (3,200) left, rho, phi, the, "bin = 0083"
    elseif  ((phi.ge.138.00000000).and.(phi.lt.144.00000000)) then
        write (3,200) left, rho, phi, the, "bin = 0084"

*****************************************************************************

          <  Program is interrupted at this stage due to space >
          <  limitations.  This program iterates over theta    >
          <  in increments of eight degrees, and each such     >
          <  time, goes over phi in increments of six degrees. >
          <  Program resumes below until end of file.          >

*****************************************************************************

    elseif  ((phi.ge.150.00000000).and.(phi.lt.156.00000000)) then
```

```
          write (3,200) left, rho, phi, the, "bin = 1286"
     elseif  ((phi.ge.156.00000000).and.(phi.lt.162.00000000)) then
          write (3,200) left, rho, phi, the, "bin = 1287"
     elseif  ((phi.ge.162.00000000).and.(phi.lt.168.00000000)) then
          write (3,200) left, rho, phi, the, "bin = 1288"
     elseif  ((phi.ge.168.00000000).and.(phi.lt.174.00000000)) then
          write (3,200) left, rho, phi, the, "bin = 1289"
     elseif  ((phi.ge.174.00000000).and.(phi.lt.180.00000000)) then
          write (3,200) left, rho, phi, the, "bin = 1290"
     endif
 elseif  ((the.ge.344.00000000).and.(the.lt.352.00000000)) then
     if  ((phi.ge.0.00000000).and.(phi.lt.6.00000000)) then
          write (3,200) left, rho, phi, the, "bin = 1291"
     elseif  ((phi.ge.6.00000000).and.(phi.lt.12.00000000)) then
          write (3,200) left, rho, phi, the, "bin = 1292"
     elseif  ((phi.ge.12.00000000).and.(phi.lt.18.00000000)) then
          write (3,200) left, rho, phi, the, "bin = 1293"
     elseif  ((phi.ge.18.00000000).and.(phi.lt.24.00000000)) then
          write (3,200) left, rho, phi, the, "bin = 1294"
     elseif  ((phi.ge.24.00000000).and.(phi.lt.30.00000000)) then
          write (3,200) left, rho, phi, the, "bin = 1295"
     elseif  ((phi.ge.30.00000000).and.(phi.lt.36.00000000)) then
          write (3,200) left, rho, phi, the, "bin = 1296"
     elseif  ((phi.ge.36.00000000).and.(phi.lt.42.00000000)) then
          write (3,200) left, rho, phi, the, "bin = 1297"
     elseif  ((phi.ge.42.00000000).and.(phi.lt.48.00000000)) then
          write (3,200) left, rho, phi, the, "bin = 1298"
     elseif  ((phi.ge.48.00000000).and.(phi.lt.54.00000000)) then
          write (3,200) left, rho, phi, the, "bin = 1299"
     elseif  ((phi.ge.54.00000000).and.(phi.lt.60.00000000)) then
          write (3,200) left, rho, phi, the, "bin = 1300"
     elseif  ((phi.ge.60.00000000).and.(phi.lt.66.00000000)) then
          write (3,200) left, rho, phi, the, "bin = 1301"
     elseif  ((phi.ge.66.00000000).and.(phi.lt.72.00000000)) then
          write (3,200) left, rho, phi, the, "bin = 1302"
     elseif  ((phi.ge.72.00000000).and.(phi.lt.78.00000000)) then
          write (3,200) left, rho, phi, the, "bin = 1303"
     elseif  ((phi.ge.78.00000000).and.(phi.lt.84.00000000)) then
          write (3,200) left, rho, phi, the, "bin = 1304"
     elseif  ((phi.ge.84.00000000).and.(phi.lt.90.00000000)) then
          write (3,200) left, rho, phi, the, "bin = 1305"
     elseif  ((phi.ge.90.00000000).and.(phi.lt.96.00000000)) then
          write (3,200) left, rho, phi, the, "bin = 1306"
     elseif  ((phi.ge.96.00000000).and.(phi.lt.102.00000000)) then
          write (3,200) left, rho, phi, the, "bin = 1307"
     elseif  ((phi.ge.102.00000000).and.(phi.lt.108.00000000)) then
          write (3,200) left, rho, phi, the, "bin = 1308"
     elseif  ((phi.ge.108.00000000).and.(phi.lt.114.00000000)) then
          write (3,200) left, rho, phi, the, "bin = 1309"
     elseif  ((phi.ge.114.00000000).and.(phi.lt.120.00000000)) then
          write (3,200) left, rho, phi, the, "bin = 1310"
     elseif  ((phi.ge.120.00000000).and.(phi.lt.126.00000000)) then
          write (3,200) left, rho, phi, the, "bin = 1311"
     elseif  ((phi.ge.126.00000000).and.(phi.lt.132.00000000)) then
          write (3,200) left, rho, phi, the, "bin = 1312"
     elseif  ((phi.ge.132.00000000).and.(phi.lt.138.00000000)) then
          write (3,200) left, rho, phi, the, "bin = 1313"
     elseif  ((phi.ge.138.00000000).and.(phi.lt.144.00000000)) then
          write (3,200) left, rho, phi, the, "bin = 1314"
     elseif  ((phi.ge.144.00000000).and.(phi.lt.150.00000000)) then
          write (3,200) left, rho, phi, the, "bin = 1315"
     elseif  ((phi.ge.150.00000000).and.(phi.lt.156.00000000)) then
```

```
          write (3,200) left, rho, phi, the, "bin = 1316"
     elseif  ((phi.ge.156.00000000).and.(phi.lt.162.00000000)) then
          write (3,200) left, rho, phi, the, "bin = 1317"
     elseif  ((phi.ge.162.00000000).and.(phi.lt.168.00000000)) then
          write (3,200) left, rho, phi, the, "bin = 1318"
     elseif  ((phi.ge.168.00000000).and.(phi.lt.174.00000000)) then
          write (3,200) left, rho, phi, the, "bin = 1319"
     elseif  ((phi.ge.174.00000000).and.(phi.lt.180.00000000)) then
          write (3,200) left, rho, phi, the, "bin = 1320"
     endif
 elseif  ((the.ge.352.00000000).and.(the.lt.360.00000000)) then
     if  ((phi.ge.0.00000000).and.(phi.lt.6.00000000)) then
          write (3,200) left, rho, phi, the, "bin = 1321"
     elseif  ((phi.ge.6.00000000).and.(phi.lt.12.00000000)) then
          write (3,200) left, rho, phi, the, "bin = 1322"
     elseif  ((phi.ge.12.00000000).and.(phi.lt.18.00000000)) then
          write (3,200) left, rho, phi, the, "bin = 1323"
     elseif  ((phi.ge.18.00000000).and.(phi.lt.24.00000000)) then
          write (3,200) left, rho, phi, the, "bin = 1324"
     elseif  ((phi.ge.24.00000000).and.(phi.lt.30.00000000)) then
          write (3,200) left, rho, phi, the, "bin = 1325"
     elseif  ((phi.ge.30.00000000).and.(phi.lt.36.00000000)) then
          write (3,200) left, rho, phi, the, "bin = 1326"
     elseif  ((phi.ge.36.00000000).and.(phi.lt.42.00000000)) then
          write (3,200) left, rho, phi, the, "bin = 1327"
     elseif  ((phi.ge.42.00000000).and.(phi.lt.48.00000000)) then
          write (3,200) left, rho, phi, the, "bin = 1328"
     elseif  ((phi.ge.48.00000000).and.(phi.lt.54.00000000)) then
          write (3,200) left, rho, phi, the, "bin = 1329"
     elseif  ((phi.ge.54.00000000).and.(phi.lt.60.00000000)) then
          write (3,200) left, rho, phi, the, "bin = 1330"
     elseif  ((phi.ge.60.00000000).and.(phi.lt.66.00000000)) then
          write (3,200) left, rho, phi, the, "bin = 1331"
     elseif  ((phi.ge.66.00000000).and.(phi.lt.72.00000000)) then
          write (3,200) left, rho, phi, the, "bin = 1332"
     elseif  ((phi.ge.72.00000000).and.(phi.lt.78.00000000)) then
          write (3,200) left, rho, phi, the, "bin = 1333"
     elseif  ((phi.ge.78.00000000).and.(phi.lt.84.00000000)) then
          write (3,200) left, rho, phi, the, "bin = 1334"
     elseif  ((phi.ge.84.00000000).and.(phi.lt.90.00000000)) then
          write (3,200) left, rho, phi, the, "bin = 1335"
     elseif  ((phi.ge.90.00000000).and.(phi.lt.96.00000000)) then
          write (3,200) left, rho, phi, the, "bin = 1336"
     elseif  ((phi.ge.96.00000000).and.(phi.lt.102.00000000)) then
          write (3,200) left, rho, phi, the, "bin = 1337"
     elseif  ((phi.ge.102.00000000).and.(phi.lt.108.00000000)) then
          write (3,200) left, rho, phi, the, "bin = 1338"
     elseif  ((phi.ge.108.00000000).and.(phi.lt.114.00000000)) then
          write (3,200) left, rho, phi, the, "bin = 1339"
     elseif  ((phi.ge.114.00000000).and.(phi.lt.120.00000000)) then
          write (3,200) left, rho, phi, the, "bin = 1340"
     elseif  ((phi.ge.120.00000000).and.(phi.lt.126.00000000)) then
          write (3,200) left, rho, phi, the, "bin = 1341"
     elseif  ((phi.ge.126.00000000).and.(phi.lt.132.00000000)) then
          write (3,200) left, rho, phi, the, "bin = 1342"
     elseif  ((phi.ge.132.00000000).and.(phi.lt.138.00000000)) then
          write (3,200) left, rho, phi, the, "bin = 1343"
     elseif  ((phi.ge.138.00000000).and.(phi.lt.144.00000000)) then
          write (3,200) left, rho, phi, the, "bin = 1344"
     elseif  ((phi.ge.144.00000000).and.(phi.lt.150.00000000)) then
          write (3,200) left, rho, phi, the, "bin = 1345"
     elseif  ((phi.ge.150.00000000).and.(phi.lt.156.00000000)) then
```

```
          write (3,200) left, rho, phi, the, "bin = 1346"
     elseif  ((phi.ge.156.00000000).and.(phi.lt.162.00000000)) then
          write (3,200) left, rho, phi, the, "bin = 1347"
     elseif  ((phi.ge.162.00000000).and.(phi.lt.168.00000000)) then
          write (3,200) left, rho, phi, the, "bin = 1348"
     elseif  ((phi.ge.168.00000000).and.(phi.lt.174.00000000)) then
          write (3,200) left, rho, phi, the, "bin = 1349"
     elseif  ((phi.ge.174.00000000).and.(phi.lt.180.00000000)) then
          write (3,200) left, rho, phi, the, "bin = 1350"
     endif
endif

!!!! body of 'if' statements end here


      go to 123
111   continue

      close(3)
      close(2)
      close(1)

      stop
      end
```

**##########    End of Pogram  "phi6_theta8_binning.f90"  ##########**

**Program 8:**

**##########    Start of Pogram   "phi10_theta10_binning.f90" ##########**

```
c     program name: find_VDWints.f

c c c c c c c c c c c c c c c c c c c c c c c c c c c c c c c c c c c c c c
c                                                                         c
c   Author:  Vicente M. Reyes, Ph.D.                                      c
c            Dept. of Pharmacol., Skaggs Sch. of Pharm. & Pharm. Sci.     c
c         &  Dept. of Integrative Biosci., S.D. Supercomptr. Ctr.         c
c            La Jolla, CA  92093-0505  U.S.A.                             c
c                                                                         c
c c c c c c c c c c c c c c c c c c c c c c c c c c c c c c c c c c c c c c

program phi10_theta10_binning

!  Bins are 10 degrees in both phi and theta.
!  Phi goes from 0 to 180 deg, so 180/10 = 18 levels in phi.
!  Theta goes from 0 to 360 deg, so 360/10 = 36 levels in theta.
!  Total no. of bins = 18x36 = 648 bins.

      character*30 left
      character*30 right
      real rho, phi, the

      open (unit =1, file = "filei")
      open (unit =3, file = "fileo")

123   read(1,200, end=111) left, rho, phi, the, right
```

```
200   format(A30, f14.8, f14.8, f14.8, A30)

!!!! body of 'if' statements start here

    if  ((the.ge.0.00000000).and.(the.lt.10.00000000)) then
        if  ((phi.ge.0.00000000).and.(phi.lt.10.00000000)) then
          write (3,200) left, rho, phi, the, "bin = 0001"
    elseif  ((phi.ge.10.00000000).and.(phi.lt.20.00000000)) then
          write (3,200) left, rho, phi, the, "bin = 0002"
    elseif  ((phi.ge.20.00000000).and.(phi.lt.30.00000000)) then
          write (3,200) left, rho, phi, the, "bin = 0003"
    elseif  ((phi.ge.30.00000000).and.(phi.lt.40.00000000)) then
          write (3,200) left, rho, phi, the, "bin = 0004"
    elseif  ((phi.ge.40.00000000).and.(phi.lt.50.00000000)) then
          write (3,200) left, rho, phi, the, "bin = 0005"
    elseif  ((phi.ge.50.00000000).and.(phi.lt.60.00000000)) then
          write (3,200) left, rho, phi, the, "bin = 0006"
    elseif  ((phi.ge.60.00000000).and.(phi.lt.70.00000000)) then
          write (3,200) left, rho, phi, the, "bin = 0007"
    elseif  ((phi.ge.70.00000000).and.(phi.lt.80.00000000)) then
          write (3,200) left, rho, phi, the, "bin = 0008"
    elseif  ((phi.ge.80.00000000).and.(phi.lt.90.00000000)) then
          write (3,200) left, rho, phi, the, "bin = 0009"
    elseif  ((phi.ge.90.00000000).and.(phi.lt.100.00000000)) then
          write (3,200) left, rho, phi, the, "bin = 0010"
    elseif  ((phi.ge.100.00000000).and.(phi.lt.110.00000000)) then
          write (3,200) left, rho, phi, the, "bin = 0011"
    elseif  ((phi.ge.110.00000000).and.(phi.lt.120.00000000)) then
          write (3,200) left, rho, phi, the, "bin = 0012"
    elseif  ((phi.ge.120.00000000).and.(phi.lt.130.00000000)) then
          write (3,200) left, rho, phi, the, "bin = 0013"
    elseif  ((phi.ge.130.00000000).and.(phi.lt.140.00000000)) then
          write (3,200) left, rho, phi, the, "bin = 0014"
    elseif  ((phi.ge.140.00000000).and.(phi.lt.150.00000000)) then
          write (3,200) left, rho, phi, the, "bin = 0015"
    elseif  ((phi.ge.150.00000000).and.(phi.lt.160.00000000)) then
          write (3,200) left, rho, phi, the, "bin = 0016"
    elseif  ((phi.ge.160.00000000).and.(phi.lt.170.00000000)) then
          write (3,200) left, rho, phi, the, "bin = 0017"
    elseif  ((phi.ge.170.00000000).and.(phi.lt.180.00000000)) then
          write (3,200) left, rho, phi, the, "bin = 0018"
     endif
elseif  ((the.ge.10.00000000).and.(the.lt.20.00000000)) then
        if  ((phi.ge.0.00000000).and.(phi.lt.10.00000000)) then
          write (3,200) left, rho, phi, the, "bin = 0019"
    elseif  ((phi.ge.10.00000000).and.(phi.lt.20.00000000)) then
          write (3,200) left, rho, phi, the, "bin = 0020"
    elseif  ((phi.ge.20.00000000).and.(phi.lt.30.00000000)) then
          write (3,200) left, rho, phi, the, "bin = 0021"
    elseif  ((phi.ge.30.00000000).and.(phi.lt.40.00000000)) then
          write (3,200) left, rho, phi, the, "bin = 0022"
    elseif  ((phi.ge.40.00000000).and.(phi.lt.50.00000000)) then
          write (3,200) left, rho, phi, the, "bin = 0023"
    elseif  ((phi.ge.50.00000000).and.(phi.lt.60.00000000)) then
          write (3,200) left, rho, phi, the, "bin = 0024"
    elseif  ((phi.ge.60.00000000).and.(phi.lt.70.00000000)) then
          write (3,200) left, rho, phi, the, "bin = 0025"
    elseif  ((phi.ge.70.00000000).and.(phi.lt.80.00000000)) then
          write (3,200) left, rho, phi, the, "bin = 0026"
    elseif  ((phi.ge.80.00000000).and.(phi.lt.90.00000000)) then
          write (3,200) left, rho, phi, the, "bin = 0027"
```

```
    elseif  ((phi.ge.90.00000000).and.(phi.lt.100.00000000)) then
          write (3,200) left, rho, phi, the, "bin = 0028"
    elseif  ((phi.ge.100.00000000).and.(phi.lt.110.00000000)) then
          write (3,200) left, rho, phi, the, "bin = 0029"
    elseif  ((phi.ge.110.00000000).and.(phi.lt.120.00000000)) then
          write (3,200) left, rho, phi, the, "bin = 0030"
    elseif  ((phi.ge.120.00000000).and.(phi.lt.130.00000000)) then
          write (3,200) left, rho, phi, the, "bin = 0031"
    elseif  ((phi.ge.130.00000000).and.(phi.lt.140.00000000)) then
          write (3,200) left, rho, phi, the, "bin = 0032"
    elseif  ((phi.ge.140.00000000).and.(phi.lt.150.00000000)) then
          write (3,200) left, rho, phi, the, "bin = 0033"
    elseif  ((phi.ge.150.00000000).and.(phi.lt.160.00000000)) then
          write (3,200) left, rho, phi, the, "bin = 0034"
    elseif  ((phi.ge.160.00000000).and.(phi.lt.170.00000000)) then
          write (3,200) left, rho, phi, the, "bin = 0035"
    elseif  ((phi.ge.170.00000000).and.(phi.lt.180.00000000)) then
          write (3,200) left, rho, phi, the, "bin = 0036"
     endif
elseif  ((the.ge.20.00000000).and.(the.lt.30.00000000)) then
        if  ((phi.ge.0.00000000).and.(phi.lt.10.00000000)) then
          write (3,200) left, rho, phi, the, "bin = 0037"
    elseif  ((phi.ge.10.00000000).and.(phi.lt.20.00000000)) then
          write (3,200) left, rho, phi, the, "bin = 0038"
    elseif  ((phi.ge.20.00000000).and.(phi.lt.30.00000000)) then
          write (3,200) left, rho, phi, the, "bin = 0039"
    elseif  ((phi.ge.30.00000000).and.(phi.lt.40.00000000)) then
          write (3,200) left, rho, phi, the, "bin = 0040"
    elseif  ((phi.ge.40.00000000).and.(phi.lt.50.00000000)) then
          write (3,200) left, rho, phi, the, "bin = 0041"
    elseif  ((phi.ge.50.00000000).and.(phi.lt.60.00000000)) then
          write (3,200) left, rho, phi, the, "bin = 0042"
    elseif  ((phi.ge.60.00000000).and.(phi.lt.70.00000000)) then
          write (3,200) left, rho, phi, the, "bin = 0043"
    elseif  ((phi.ge.70.00000000).and.(phi.lt.80.00000000)) then
          write (3,200) left, rho, phi, the, "bin = 0044"
    elseif  ((phi.ge.80.00000000).and.(phi.lt.90.00000000)) then
          write (3,200) left, rho, phi, the, "bin = 0045"
    elseif  ((phi.ge.90.00000000).and.(phi.lt.100.00000000)) then
          write (3,200) left, rho, phi, the, "bin = 0046"
    elseif  ((phi.ge.100.00000000).and.(phi.lt.110.00000000)) then
          write (3,200) left, rho, phi, the, "bin = 0047"
    elseif  ((phi.ge.110.00000000).and.(phi.lt.120.00000000)) then
          write (3,200) left, rho, phi, the, "bin = 0048"


*******************************************************************************


           <  Program is interrupted at this stage due to space >
           <  limitations.  This program iterates over theta    >
           <  in increments of ten degrees, and each such time  >
           <  goes over phi in increments of six degrees.       >
           <  Program resumes below until end of file.          >


**********************************************************************************


    elseif  ((phi.ge.120.00000000).and.(phi.lt.130.00000000)) then
          write (3,200) left, rho, phi, the, "bin = 0571"
```

```
    elseif  ((phi.ge.130.00000000).and.(phi.lt.140.00000000)) then
          write (3,200) left, rho, phi, the, "bin = 0572"
    elseif  ((phi.ge.140.00000000).and.(phi.lt.150.00000000)) then
          write (3,200) left, rho, phi, the, "bin = 0573"
    elseif  ((phi.ge.150.00000000).and.(phi.lt.160.00000000)) then
          write (3,200) left, rho, phi, the, "bin = 0574"
    elseif  ((phi.ge.160.00000000).and.(phi.lt.170.00000000)) then
          write (3,200) left, rho, phi, the, "bin = 0575"
    elseif  ((phi.ge.170.00000000).and.(phi.lt.180.00000000)) then
          write (3,200) left, rho, phi, the, "bin = 0576"
     endif
elseif  ((the.ge.320.00000000).and.(the.lt.330.00000000)) then
        if  ((phi.ge.0.00000000).and.(phi.lt.10.00000000)) then
          write (3,200) left, rho, phi, the, "bin = 0577"
    elseif  ((phi.ge.10.00000000).and.(phi.lt.20.00000000)) then
          write (3,200) left, rho, phi, the, "bin = 0578"
    elseif  ((phi.ge.20.00000000).and.(phi.lt.30.00000000)) then
          write (3,200) left, rho, phi, the, "bin = 0579"
    elseif  ((phi.ge.30.00000000).and.(phi.lt.40.00000000)) then
          write (3,200) left, rho, phi, the, "bin = 0580"
    elseif  ((phi.ge.40.00000000).and.(phi.lt.50.00000000)) then
          write (3,200) left, rho, phi, the, "bin = 0581"
    elseif  ((phi.ge.50.00000000).and.(phi.lt.60.00000000)) then
          write (3,200) left, rho, phi, the, "bin = 0582"
    elseif  ((phi.ge.60.00000000).and.(phi.lt.70.00000000)) then
          write (3,200) left, rho, phi, the, "bin = 0583"
    elseif  ((phi.ge.70.00000000).and.(phi.lt.80.00000000)) then
          write (3,200) left, rho, phi, the, "bin = 0584"
    elseif  ((phi.ge.80.00000000).and.(phi.lt.90.00000000)) then
          write (3,200) left, rho, phi, the, "bin = 0585"
    elseif  ((phi.ge.90.00000000).and.(phi.lt.100.00000000)) then
          write (3,200) left, rho, phi, the, "bin = 0586"
    elseif  ((phi.ge.100.00000000).and.(phi.lt.110.00000000)) then
          write (3,200) left, rho, phi, the, "bin = 0587"
    elseif  ((phi.ge.110.00000000).and.(phi.lt.120.00000000)) then
          write (3,200) left, rho, phi, the, "bin = 0588"
    elseif  ((phi.ge.120.00000000).and.(phi.lt.130.00000000)) then
          write (3,200) left, rho, phi, the, "bin = 0589"
    elseif  ((phi.ge.130.00000000).and.(phi.lt.140.00000000)) then
          write (3,200) left, rho, phi, the, "bin = 0590"
    elseif  ((phi.ge.140.00000000).and.(phi.lt.150.00000000)) then
          write (3,200) left, rho, phi, the, "bin = 0591"
    elseif  ((phi.ge.150.00000000).and.(phi.lt.160.00000000)) then
          write (3,200) left, rho, phi, the, "bin = 0592"
    elseif  ((phi.ge.160.00000000).and.(phi.lt.170.00000000)) then
          write (3,200) left, rho, phi, the, "bin = 0593"
    elseif  ((phi.ge.170.00000000).and.(phi.lt.180.00000000)) then
          write (3,200) left, rho, phi, the, "bin = 0594"
     endif
elseif  ((the.ge.330.00000000).and.(the.lt.340.00000000)) then
        if  ((phi.ge.0.00000000).and.(phi.lt.10.00000000)) then
          write (3,200) left, rho, phi, the, "bin = 0595"
    elseif  ((phi.ge.10.00000000).and.(phi.lt.20.00000000)) then
          write (3,200) left, rho, phi, the, "bin = 0596"
    elseif  ((phi.ge.20.00000000).and.(phi.lt.30.00000000)) then
          write (3,200) left, rho, phi, the, "bin = 0597"
    elseif  ((phi.ge.30.00000000).and.(phi.lt.40.00000000)) then
          write (3,200) left, rho, phi, the, "bin = 0598"
    elseif  ((phi.ge.40.00000000).and.(phi.lt.50.00000000)) then
          write (3,200) left, rho, phi, the, "bin = 0599"
    elseif  ((phi.ge.50.00000000).and.(phi.lt.60.00000000)) then
          write (3,200) left, rho, phi, the, "bin = 0600"
```

```
    elseif  ((phi.ge.60.00000000).and.(phi.lt.70.00000000)) then
          write (3,200) left, rho, phi, the, "bin = 0601"
    elseif  ((phi.ge.70.00000000).and.(phi.lt.80.00000000)) then
          write (3,200) left, rho, phi, the, "bin = 0602"
    elseif  ((phi.ge.80.00000000).and.(phi.lt.90.00000000)) then
          write (3,200) left, rho, phi, the, "bin = 0603"
    elseif  ((phi.ge.90.00000000).and.(phi.lt.100.00000000)) then
          write (3,200) left, rho, phi, the, "bin = 0604"
    elseif  ((phi.ge.100.00000000).and.(phi.lt.110.00000000)) then
          write (3,200) left, rho, phi, the, "bin = 0605"
    elseif  ((phi.ge.110.00000000).and.(phi.lt.120.00000000)) then
          write (3,200) left, rho, phi, the, "bin = 0606"
    elseif  ((phi.ge.120.00000000).and.(phi.lt.130.00000000)) then
          write (3,200) left, rho, phi, the, "bin = 0607"
    elseif  ((phi.ge.130.00000000).and.(phi.lt.140.00000000)) then
          write (3,200) left, rho, phi, the, "bin = 0608"
    elseif  ((phi.ge.140.00000000).and.(phi.lt.150.00000000)) then
          write (3,200) left, rho, phi, the, "bin = 0609"
    elseif  ((phi.ge.150.00000000).and.(phi.lt.160.00000000)) then
          write (3,200) left, rho, phi, the, "bin = 0610"
    elseif  ((phi.ge.160.00000000).and.(phi.lt.170.00000000)) then
          write (3,200) left, rho, phi, the, "bin = 0611"
    elseif  ((phi.ge.170.00000000).and.(phi.lt.180.00000000)) then
          write (3,200) left, rho, phi, the, "bin = 0612"
     endif
elseif  ((the.ge.340.00000000).and.(the.lt.350.00000000)) then
        if  ((phi.ge.0.00000000).and.(phi.lt.10.00000000)) then
          write (3,200) left, rho, phi, the, "bin = 0613"
    elseif  ((phi.ge.10.00000000).and.(phi.lt.20.00000000)) then
          write (3,200) left, rho, phi, the, "bin = 0614"
    elseif  ((phi.ge.20.00000000).and.(phi.lt.30.00000000)) then
          write (3,200) left, rho, phi, the, "bin = 0615"
    elseif  ((phi.ge.30.00000000).and.(phi.lt.40.00000000)) then
          write (3,200) left, rho, phi, the, "bin = 0616"
    elseif  ((phi.ge.40.00000000).and.(phi.lt.50.00000000)) then
          write (3,200) left, rho, phi, the, "bin = 0617"
    elseif  ((phi.ge.50.00000000).and.(phi.lt.60.00000000)) then
          write (3,200) left, rho, phi, the, "bin = 0618"
    elseif  ((phi.ge.60.00000000).and.(phi.lt.70.00000000)) then
          write (3,200) left, rho, phi, the, "bin = 0619"
    elseif  ((phi.ge.70.00000000).and.(phi.lt.80.00000000)) then
          write (3,200) left, rho, phi, the, "bin = 0620"
    elseif  ((phi.ge.80.00000000).and.(phi.lt.90.00000000)) then
          write (3,200) left, rho, phi, the, "bin = 0621"
    elseif  ((phi.ge.90.00000000).and.(phi.lt.100.00000000)) then
          write (3,200) left, rho, phi, the, "bin = 0622"
    elseif  ((phi.ge.100.00000000).and.(phi.lt.110.00000000)) then
          write (3,200) left, rho, phi, the, "bin = 0623"
    elseif  ((phi.ge.110.00000000).and.(phi.lt.120.00000000)) then
          write (3,200) left, rho, phi, the, "bin = 0624"
    elseif  ((phi.ge.120.00000000).and.(phi.lt.130.00000000)) then
          write (3,200) left, rho, phi, the, "bin = 0625"
    elseif  ((phi.ge.130.00000000).and.(phi.lt.140.00000000)) then
          write (3,200) left, rho, phi, the, "bin = 0626"
    elseif  ((phi.ge.140.00000000).and.(phi.lt.150.00000000)) then
          write (3,200) left, rho, phi, the, "bin = 0627"
    elseif  ((phi.ge.150.00000000).and.(phi.lt.160.00000000)) then
          write (3,200) left, rho, phi, the, "bin = 0628"
    elseif  ((phi.ge.160.00000000).and.(phi.lt.170.00000000)) then
          write (3,200) left, rho, phi, the, "bin = 0629"
    elseif  ((phi.ge.170.00000000).and.(phi.lt.180.00000000)) then
          write (3,200) left, rho, phi, the, "bin = 0630"
```

```
     endif
elseif  ((the.ge.350.00000000).and.(the.lt.360.00000000)) then
        if  ((phi.ge.0.00000000).and.(phi.lt.10.00000000)) then
          write (3,200) left, rho, phi, the, "bin = 0631"
    elseif  ((phi.ge.10.00000000).and.(phi.lt.20.00000000)) then
          write (3,200) left, rho, phi, the, "bin = 0632"
    elseif  ((phi.ge.20.00000000).and.(phi.lt.30.00000000)) then
          write (3,200) left, rho, phi, the, "bin = 0633"
    elseif  ((phi.ge.30.00000000).and.(phi.lt.40.00000000)) then
          write (3,200) left, rho, phi, the, "bin = 0634"
    elseif  ((phi.ge.40.00000000).and.(phi.lt.50.00000000)) then
          write (3,200) left, rho, phi, the, "bin = 0635"
    elseif  ((phi.ge.50.00000000).and.(phi.lt.60.00000000)) then
          write (3,200) left, rho, phi, the, "bin = 0636"
    elseif  ((phi.ge.60.00000000).and.(phi.lt.70.00000000)) then
          write (3,200) left, rho, phi, the, "bin = 0637"
    elseif  ((phi.ge.70.00000000).and.(phi.lt.80.00000000)) then
          write (3,200) left, rho, phi, the, "bin = 0638"
    elseif  ((phi.ge.80.00000000).and.(phi.lt.90.00000000)) then
          write (3,200) left, rho, phi, the, "bin = 0639"
    elseif  ((phi.ge.90.00000000).and.(phi.lt.100.00000000)) then
          write (3,200) left, rho, phi, the, "bin = 0640"
    elseif  ((phi.ge.100.00000000).and.(phi.lt.110.00000000)) then
          write (3,200) left, rho, phi, the, "bin = 0641"
    elseif  ((phi.ge.110.00000000).and.(phi.lt.120.00000000)) then
          write (3,200) left, rho, phi, the, "bin = 0642"
    elseif  ((phi.ge.120.00000000).and.(phi.lt.130.00000000)) then
          write (3,200) left, rho, phi, the, "bin = 0643"
    elseif  ((phi.ge.130.00000000).and.(phi.lt.140.00000000)) then
          write (3,200) left, rho, phi, the, "bin = 0644"
    elseif  ((phi.ge.140.00000000).and.(phi.lt.150.00000000)) then
          write (3,200) left, rho, phi, the, "bin = 0645"
    elseif  ((phi.ge.150.00000000).and.(phi.lt.160.00000000)) then
          write (3,200) left, rho, phi, the, "bin = 0646"
    elseif  ((phi.ge.160.00000000).and.(phi.lt.170.00000000)) then
          write (3,200) left, rho, phi, the, "bin = 0647"
    elseif  ((phi.ge.170.00000000).and.(phi.lt.180.00000000)) then
          write (3,200) left, rho, phi, the, "bin = 0648"
     endif
 endif

!!!! body of 'if' statements end here


      go to 123
111   continue

      close(3)
      close(2)
      close(1)

      stop
      end
```

**#########   End of Pogram  "phi10_theta10_binning.f90"  #########**

**Program 9:**

**################# Start of Pogram "bin_rho.f90" #####################**

```
program phi_theta_binning

!  Bins are 6 degrees in phi and 8 degrees in theta.
!  Phi goes from 0 to 180 deg, so 180/6 = 30 levels in phi.
!  Theta goes from 0 to 360 deg, so 360/8 = 45 levels in theta.
!  Total no. of bins = 30x45 = 1350 bins.

      character*30 left
      character*30 right
      real rho, phi, the

      open (unit =1, file = "filei")
      open (unit =3, file = "fileo")

123   read(1,100, end=111) left, rho, phi, the, right
100   format(A30, f14.8, f14.8, f14.8, A30)

    if  ((rho.ge.0.00000000).and.(rho.lt.1.50000000)) then
         write (3,200) left, rho, rho, the, right, "rhobin = 01"
    elseif  ((rho.ge.1.50000000).and.(rho.lt.3.00000000)) then
         write (3,200) left, rho, rho, the, right, "rhobin = 02"
    elseif  ((rho.ge.3.00000000).and.(rho.lt.4.50000000)) then
         write (3,200) left, rho, rho, the, right, "rhobin = 03"
    elseif  ((rho.ge.4.50000000).and.(rho.lt.6.00000000)) then
         write (3,200) left, rho, rho, the, right, "rhobin = 04"
    elseif  ((rho.ge.6.00000000).and.(rho.lt.7.50000000)) then
         write (3,200) left, rho, rho, the, right, "rhobin = 05"
    elseif  ((rho.ge.7.50000000).and.(rho.lt.9.00000000)) then
         write (3,200) left, rho, rho, the, right, "rhobin = 06"
    elseif  ((rho.ge.9.00000000).and.(rho.lt.10.50000000)) then
         write (3,200) left, rho, rho, the, right, "rhobin = 07"
    elseif  ((rho.ge.10.50000000).and.(rho.lt.12.00000000)) then
         write (3,200) left, rho, rho, the, right, "rhobin = 08"
    elseif  ((rho.ge.12.00000000).and.(rho.lt.13.50000000)) then
         write (3,200) left, rho, rho, the, right, "rhobin = 09"
    elseif  ((rho.ge.13.50000000).and.(rho.lt.15.00000000)) then
         write (3,200) left, rho, rho, the, right, "rhobin = 10"
    elseif  ((rho.ge.15.00000000).and.(rho.lt.16.50000000)) then
         write (3,200) left, rho, rho, the, right, "rhobin = 11"
    elseif  ((rho.ge.16.50000000).and.(rho.lt.18.00000000)) then
         write (3,200) left, rho, rho, the, right, "rhobin = 12"
    elseif  ((rho.ge.18.00000000).and.(rho.lt.19.50000000)) then
         write (3,200) left, rho, rho, the, right, "rhobin = 13"
    elseif  ((rho.ge.19.50000000).and.(rho.lt.21.00000000)) then
         write (3,200) left, rho, rho, the, right, "rhobin = 14"
    elseif  ((rho.ge.21.00000000).and.(rho.lt.22.50000000)) then
         write (3,200) left, rho, rho, the, right, "rhobin = 15"
    elseif  ((rho.ge.22.50000000).and.(rho.lt.24.00000000)) then
         write (3,200) left, rho, rho, the, right, "rhobin = 16"
    elseif  ((rho.ge.24.00000000).and.(rho.lt.25.50000000)) then
         write (3,200) left, rho, rho, the, right, "rhobin = 17"
    elseif  ((rho.ge.25.50000000).and.(rho.lt.27.00000000)) then
         write (3,200) left, rho, rho, the, right, "rhobin = 18"
    elseif  ((rho.ge.27.00000000).and.(rho.lt.28.50000000)) then
         write (3,200) left, rho, rho, the, right, "rhobin = 19"
    elseif  ((rho.ge.28.50000000).and.(rho.lt.30.00000000)) then
         write (3,200) left, rho, rho, the, right, "rhobin = 20"
    elseif  ((rho.ge.30.00000000).and.(rho.lt.31.50000000)) then
         write (3,200) left, rho, rho, the, right, "rhobin = 21"
```

```
    elseif  ((rho.ge.31.50000000).and.(rho.lt.33.00000000)) then
         write (3,200) left, rho, rho, the, right, "rhobin = 22"
    elseif  ((rho.ge.33.00000000).and.(rho.lt.34.50000000)) then
         write (3,200) left, rho, rho, the, right, "rhobin = 23"
    elseif  ((rho.ge.34.50000000).and.(rho.lt.36.00000000)) then
         write (3,200) left, rho, rho, the, right, "rhobin = 24"
    elseif  ((rho.ge.36.00000000).and.(rho.lt.37.50000000)) then
         write (3,200) left, rho, rho, the, right, "rhobin = 25"
    elseif  ((rho.ge.37.50000000).and.(rho.lt.39.00000000)) then
         write (3,200) left, rho, rho, the, right, "rhobin = 26"
    elseif  ((rho.ge.39.00000000).and.(rho.lt.40.50000000)) then
         write (3,200) left, rho, rho, the, right, "rhobin = 27"
    elseif  ((rho.ge.40.50000000).and.(rho.lt.42.00000000)) then
         write (3,200) left, rho, rho, the, right, "rhobin = 28"
    elseif  ((rho.ge.42.00000000).and.(rho.lt.43.50000000)) then
         write (3,200) left, rho, rho, the, right, "rhobin = 29"
    elseif  ((rho.ge.43.50000000).and.(rho.lt.45.00000000)) then
         write (3,200) left, rho, rho, the, right, "rhobin = 30"
    elseif  ((rho.ge.45.00000000).and.(rho.lt.46.50000000)) then
         write (3,200) left, rho, rho, the, right, "rhobin = 31"
    elseif  ((rho.ge.46.50000000).and.(rho.lt.48.00000000)) then
         write (3,200) left, rho, rho, the, right, "rhobin = 32"
    elseif  ((rho.ge.48.00000000).and.(rho.lt.49.50000000)) then
         write (3,200) left, rho, rho, the, right, "rhobin = 33"
    elseif  ((rho.ge.49.50000000).and.(rho.lt.51.00000000)) then
         write (3,200) left, rho, rho, the, right, "rhobin = 34"
    elseif  ((rho.ge.51.00000000).and.(rho.lt.52.50000000)) then
         write (3,200) left, rho, rho, the, right, "rhobin = 35"
    endif

200   format(A30, f14.8, f14.8, f14.8, A30, A12)

      go to 123
111   continue

      close(3)
      close(1)

      stop
      end
```

**################# End of Pogram "bin_rho.f90" ###########################**

**10. FIGURES:**

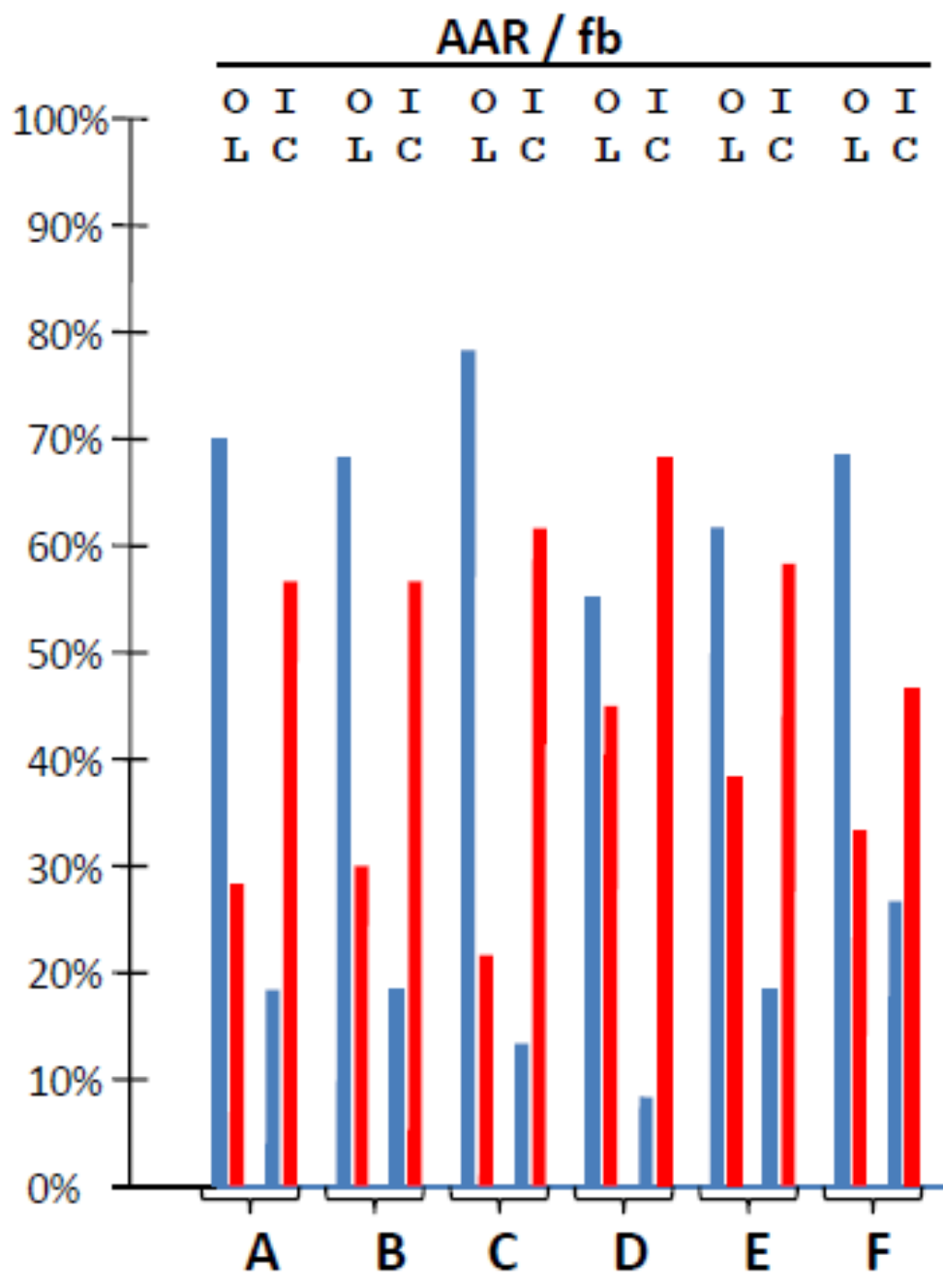

Figure 1, Panel A

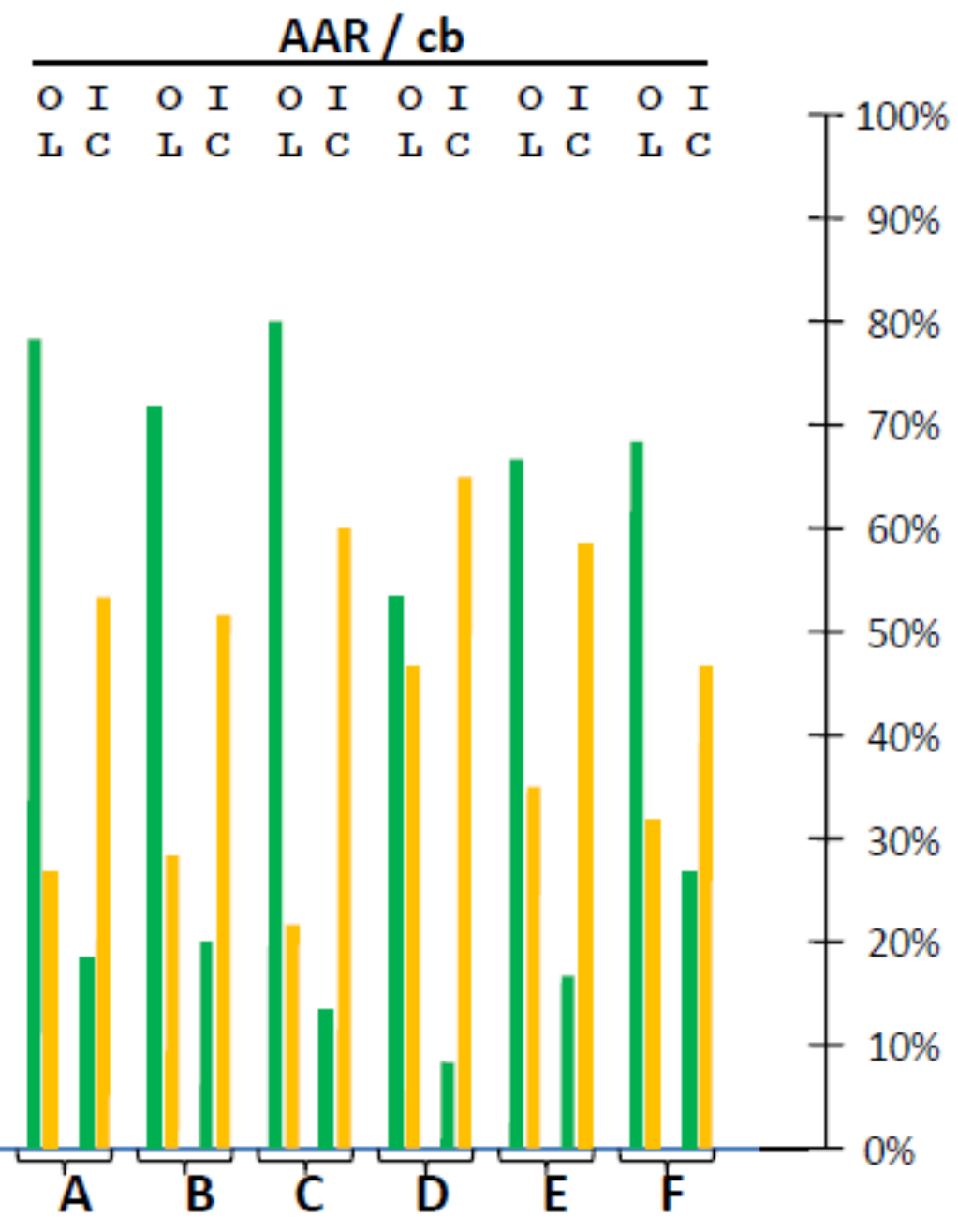

Figure 1, Panel B

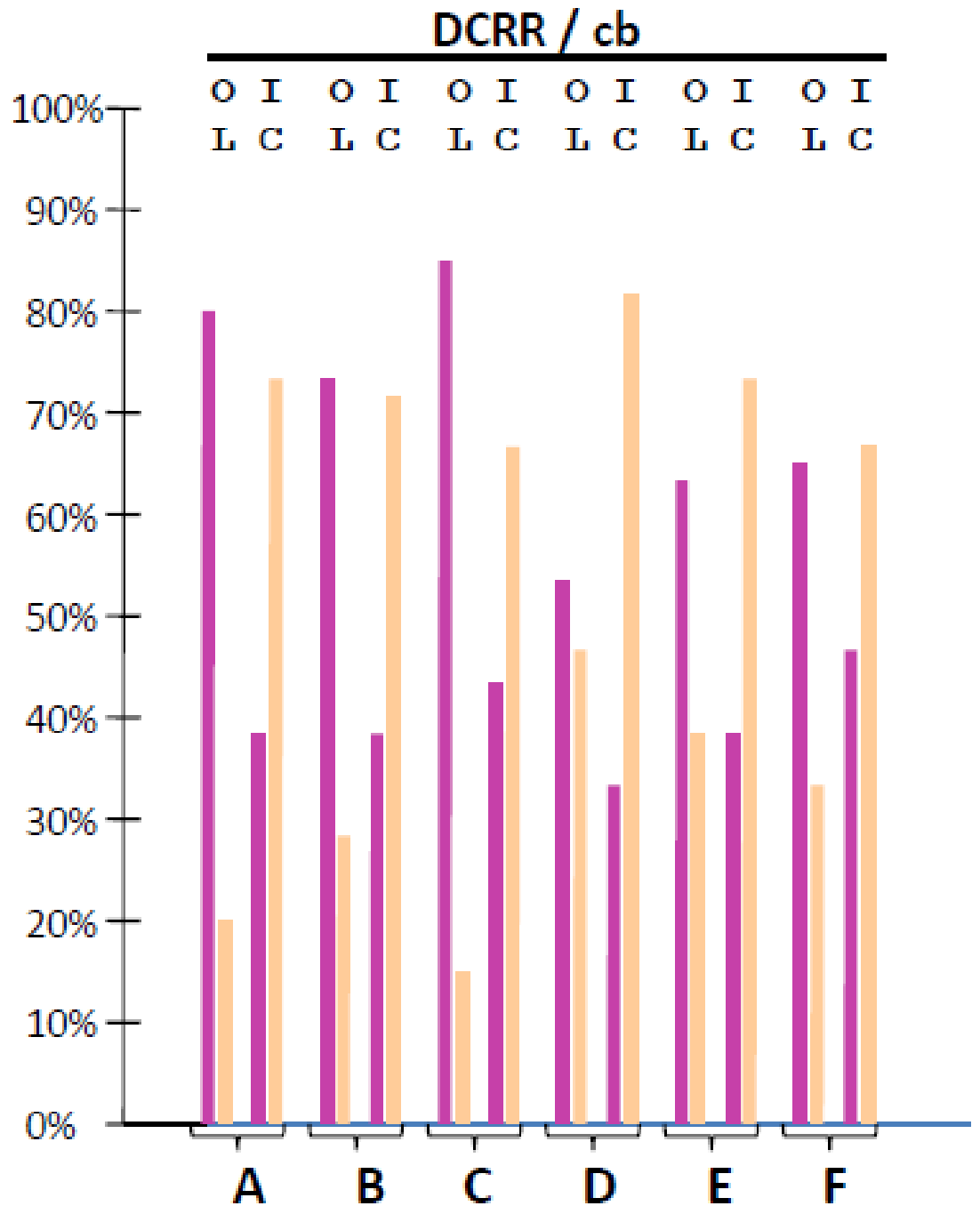

Figure 1, Panel C

Figure 1, Panel A, shows the percentage hydrophilic residues in blue, and the percentage hydropbobic residues in red, for proteins A, B, C, D and E in all-atom representation (AAR) and when fine bining (fb) is used. The actual numerical figures associated with this plot are shown in Table 3, Panel A. Figure 1, Panel B shows percentage hydrophilic residues in green, and the percentage hydropbobic residues in gold, where the proteins are in AAR and coarse binning (cb) is used. The actual numerical figures associated with this plot are shown in Table 3, Panel B. Figure 1, Panel C shows the percentage hydrophilic residues in heliotrope; and the percentage hydropbobic residues in peach, where the proteins are in double-centroid reduced representation (DCRR) and coarse binning (cb) is used. The actual numerical figures associated with this plot are shown in Table 3, Panel C. Note that in every case, there is a significantly higher % of hydrophilic residues than hydrophobic ones in the OL, while exactly the reverse is true for the IC.

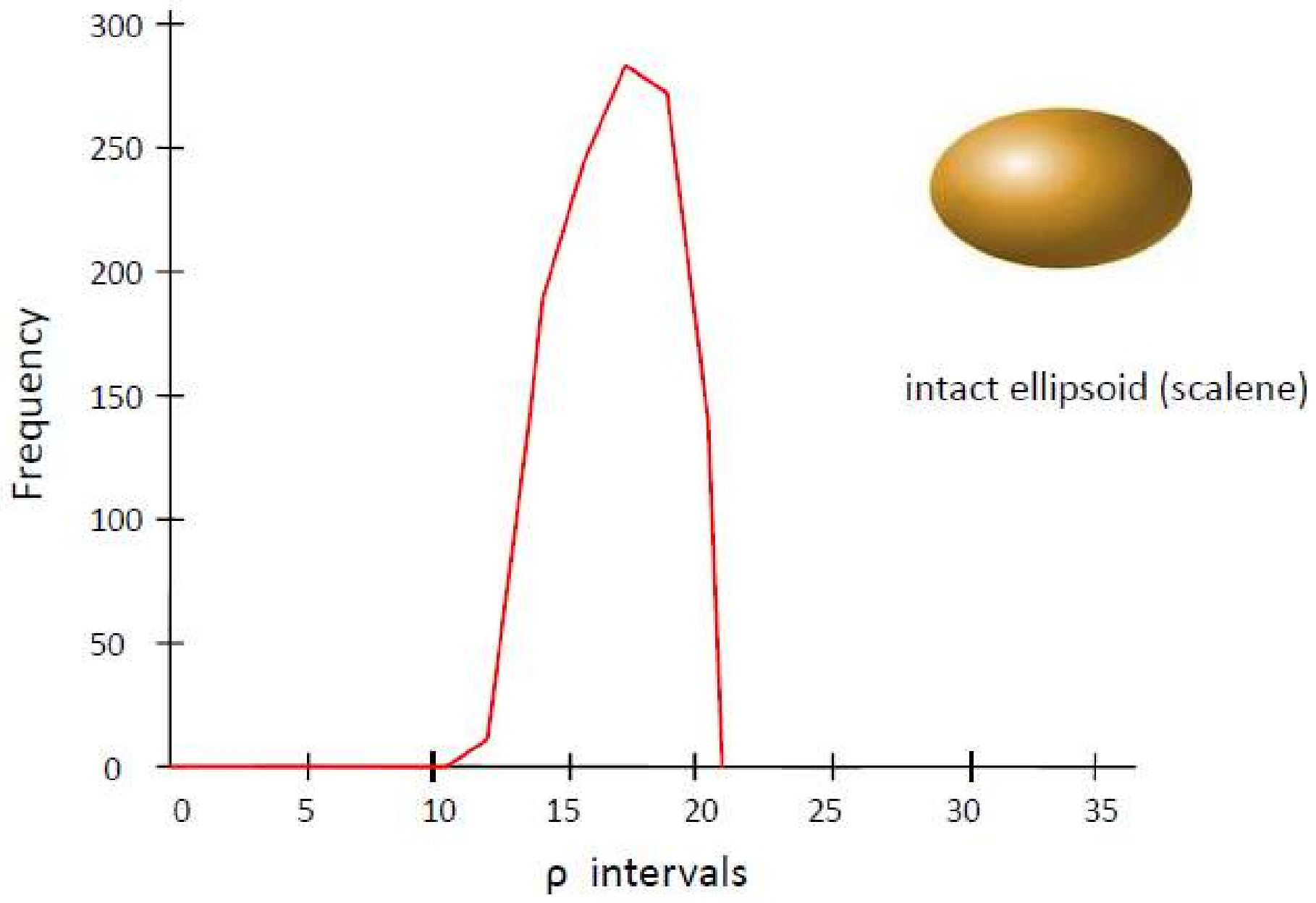

**Figure 2, Panel A**

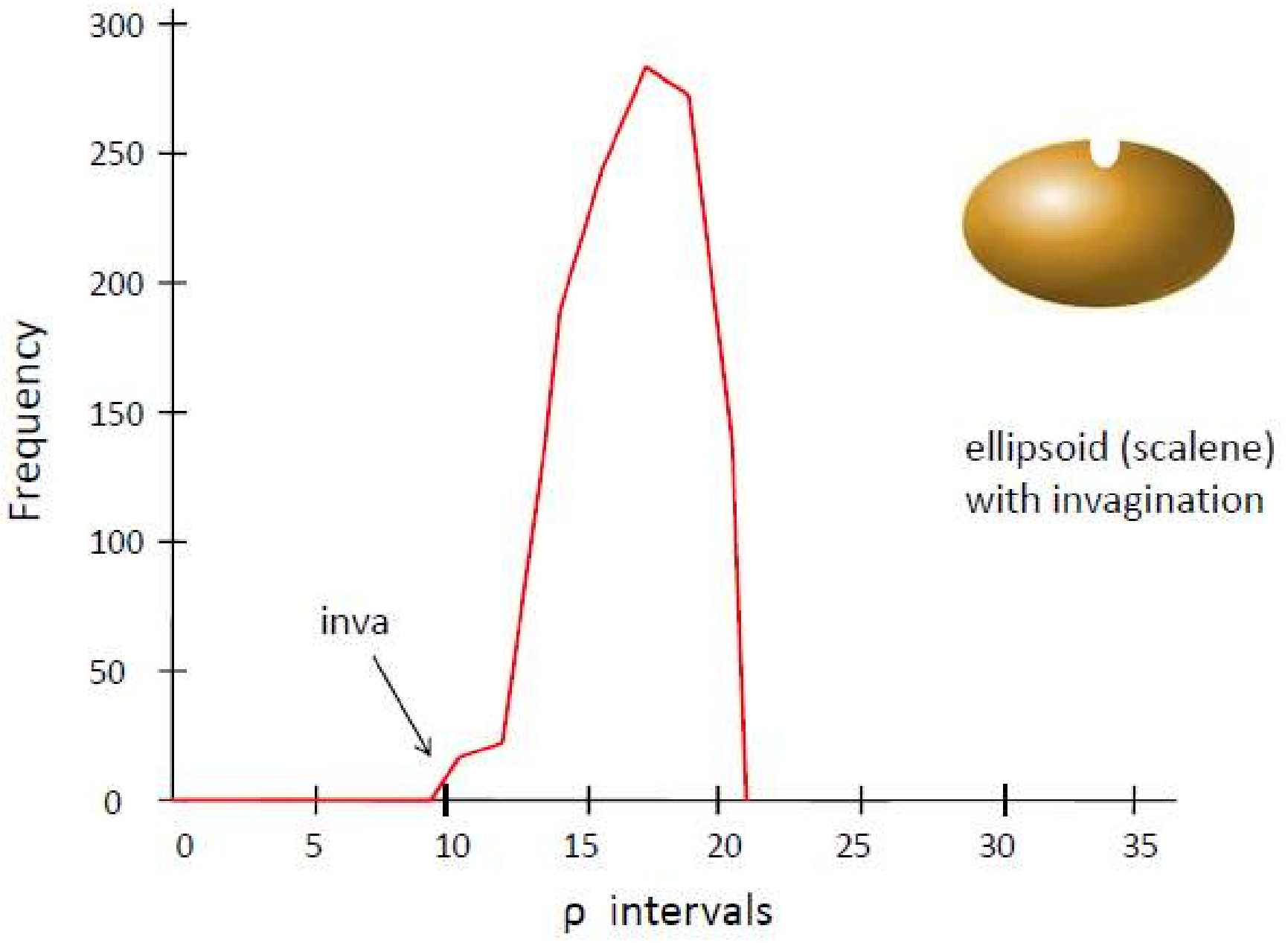

**Figure 2, Panel B**

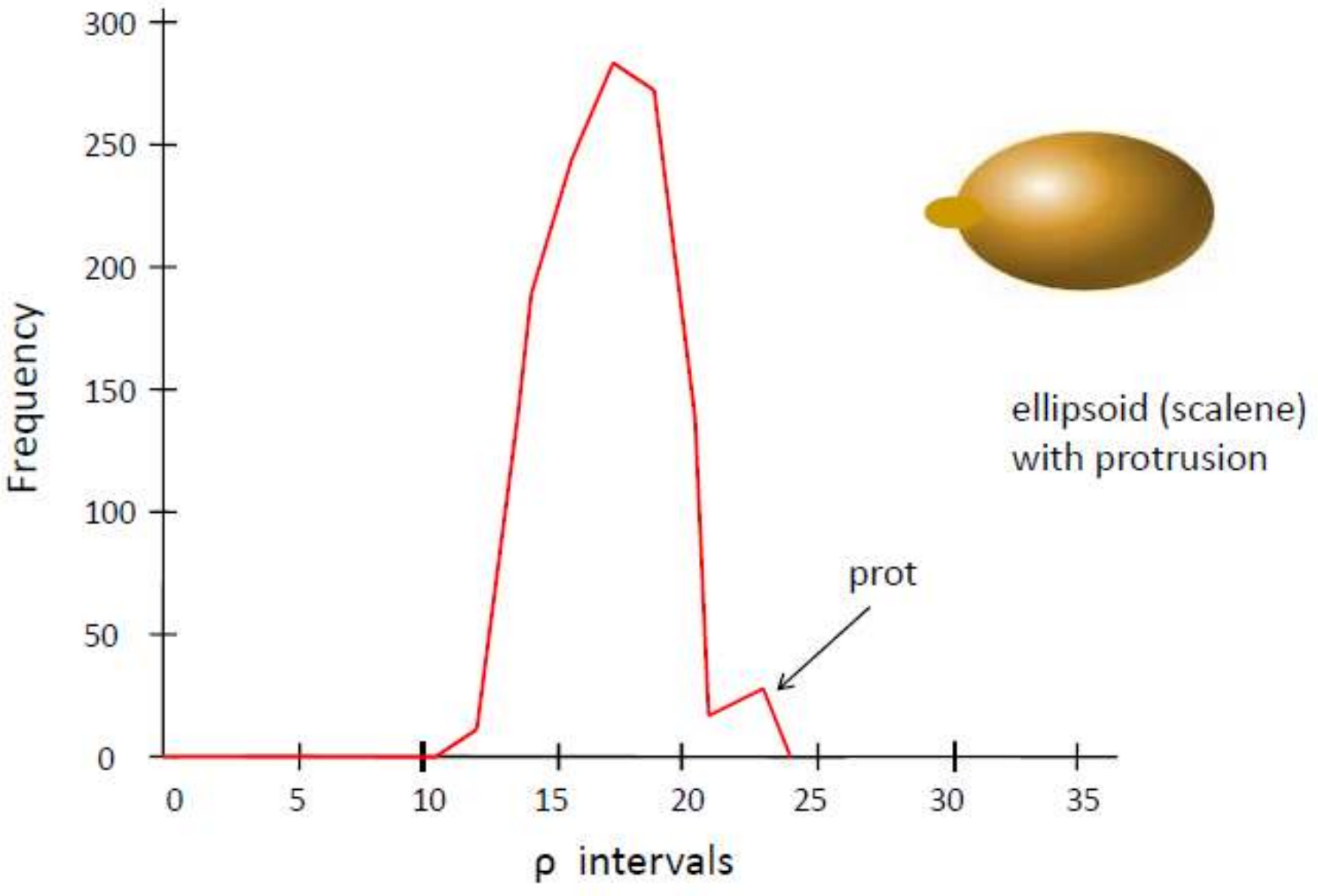

**Figure 2, Panel C**

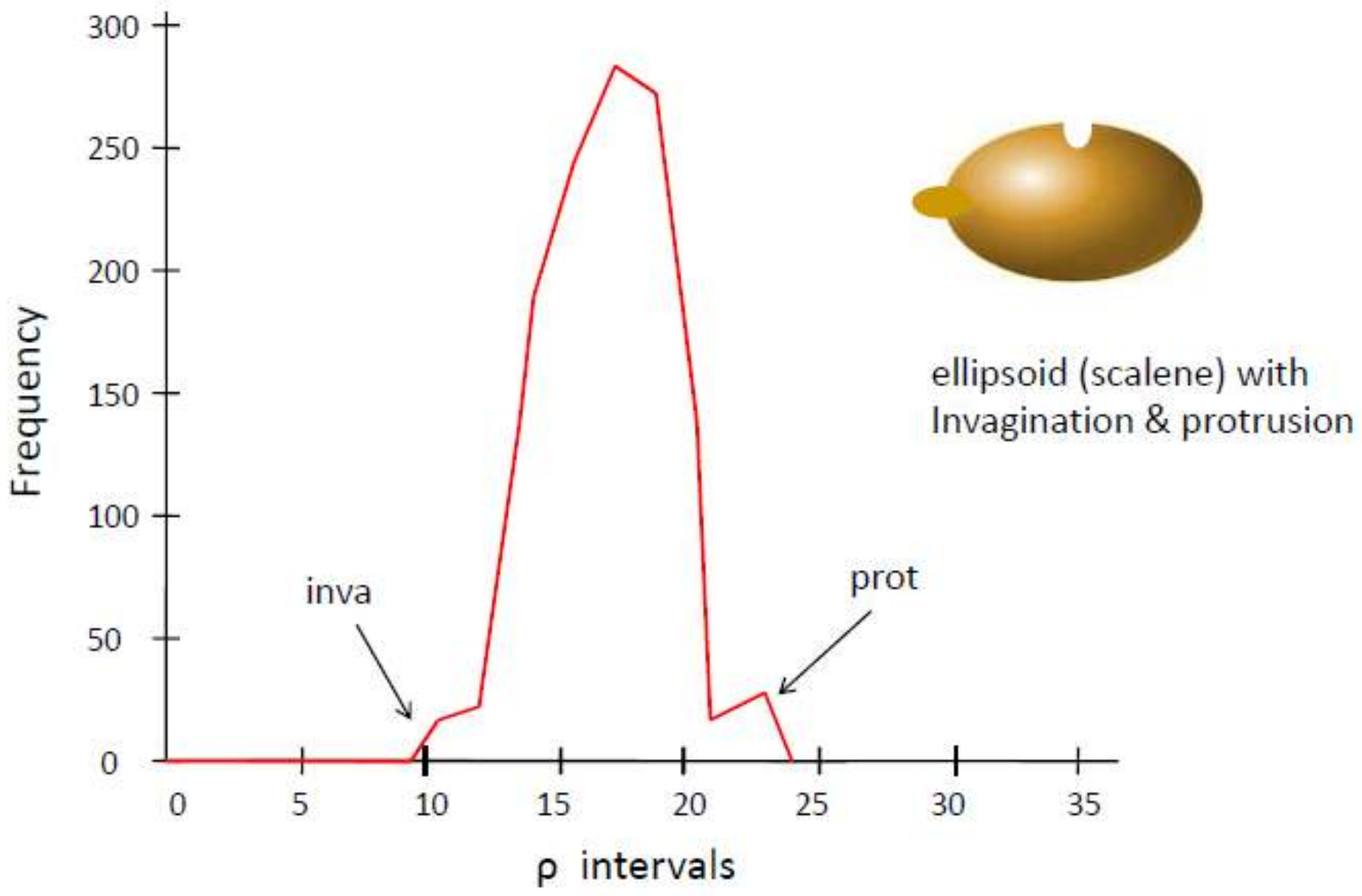

**Figure 2, Panel D**

Figure 2, Panel A, shows the plot of frequency distribution of maximum ρ's (FDMR) for an intact ellipsoid "theoretical protein" (a.k.a. "model protein") shown schematically on the right side of the plot. The model protein has no invagination or protrusion, and the FDMR plot is "clean" single peak. Figure 2, Panel B, shows the FDMR plot for an ellipsoid model protein with a single invagination shown schematically on the right side of the plot. The FDMR plot has a small subpeak on the left hand side (lagging side) of the main peak, which is taken to be indicative of an invagination on the protein ('inva'). Figure 2, Panel C, shows the FDMR plot for an ellipsoid model protein with a single protrusion shown schematically on the right side of the plot. The FDMR plot has a small subpeak on the right hand side (leading side) of the main peak, which is taken to be indicative of a protrusion on the protein ('prot'). Lastly, Figure 2, Panel D, shows the FDMR plot for an ellipsoid model protein with both an invagination and a protrusion shown schematically on the right side of the plot. The FDMR plot has two small subpeaks on the left and right hand sides (lagging and leading sides) of the main peak, which are taken to be indicative of an invagination ('inva') and a protrusion ('prot') on the protein.

**11. TABLES:**

**Protein 3D Structure Files in Spherical and Cartesian Coordinate Representation**

Spherical coordinate representation

| | | | | | | | | |
|---|---|---|---|---|---|---|---|---|
| ATOM | 1 | N | ALA | 1 | 20.14196014 | 41.89954376 | 92.45415497 | (RHO,PHI,THETA) |
| ATOM | 2 | CA | ALA | 1 | 20.16541481 | 37.91946030 | 94.28984833 | (RHO,PHI,THETA) |
| ATOM | 3 | C | ALA | 1 | 18.74901581 | 36.28638077 | 93.08413696 | (RHO,PHI,THETA) |
| ATOM | 4 | O | ALA | 1 | 18.00141144 | 38.78060913 | 90.05233765 | (RHO,PHI,THETA) |
| ATOM | 5 | CB | ALA | 1 | 20.57125282 | 37.86121368 | 101.13373566 | (RHO,PHI,THETA) |
| ATOM | 6 | N | SER | 2 | 18.38153076 | 32.48158264 | 95.31267548 | (RHO,PHI,THETA) |
| ATOM | 7 | CA | SER | 2 | 17.01415062 | 30.88849449 | 94.55062103 | (RHO,PHI,THETA) |
| ATOM | 8 | C | SER | 2 | 16.18741989 | 33.82222748 | 100.88840485 | (RHO,PHI,THETA) |
| ATOM | 9 | O | SER | 2 | 15.15668678 | 32.01140594 | 103.69378662 | (RHO,PHI,THETA) |
| ATOM | 10 | CB | SER | 2 | 16.99642754 | 25.61505890 | 97.13759613 | (RHO,PHI,THETA) |
| ATOM | 11 | OG | SER | 2 | 15.79344749 | 22.61577797 | 97.08406830 | (RHO,PHI,THETA) |
| ATOM | 12 | N | ARG | 3 | 16.56322479 | 38.12343597 | 103.54563904 | (RHO,PHI,THETA) |
| ATOM | 13 | CA | ARG | 3 | 15.89363766 | 41.08279800 | 108.95711517 | (RHO,PHI,THETA) |
| ATOM | 14 | C | ARG | 3 | 16.07167816 | 46.15010452 | 106.56809235 | (RHO,PHI,THETA) |
| ATOM | 15 | O | ARG | 3 | 17.11002350 | 47.21485901 | 103.79389191 | (RHO,PHI,THETA) |
| ATOM | 16 | CB | ARG | 3 | 16.60504150 | 41.04722977 | 116.17496490 | (RHO,PHI,THETA) |
| ATOM | 17 | CG | ARG | 3 | 16.42333984 | 36.73987198 | 121.12377930 | (RHO,PHI,THETA) |
| ATOM | 18 | CD | ARG | 3 | 16.96395874 | 37.38563156 | 129.20018005 | (RHO,PHI,THETA) |
| ATOM | 19 | NE | ARG | 3 | 18.42322350 | 37.46998215 | 129.20271301 | (RHO,PHI,THETA) |
| ATOM | 20 | CZ | ARG | 3 | 19.20050621 | 39.09282303 | 134.02285767 | (RHO,PHI,THETA) |
| ATOM | 21 | NH1 | ARG | 3 | 18.71472740 | 41.20111084 | 139.00950623 | (RHO,PHI,THETA) |
| ATOM | 22 | NH2 | ARG | 3 | 20.54468346 | 38.83226395 | 133.36126709 | (RHO,PHI,THETA) |

Cartesian coordinate representation

| | | | | | | | | | | | |
|---|---|---|---|---|---|---|---|---|---|---|---|
| ATOM | 1 | N | ALA | 1 | 14.419 | 46.428 | 83.486 | 1.00 | 47.38 | 1ADS | 70 |
| ATOM | 2 | CA | ALA | 1 | 14.068 | 45.347 | 84.402 | 1.00 | 48.01 | 1ADS | 71 |
| ATOM | 3 | C | ALA | 1 | 14.398 | 44.069 | 83.607 | 1.00 | 46.68 | 1ADS | 72 |
| ATOM | 4 | O | ALA | 1 | 14.985 | 44.264 | 82.527 | 1.00 | 47.75 | 1ADS | 73 |
| ATOM | 5 | CB | ALA | 1 | 12.557 | 45.377 | 84.735 | 1.00 | 48.62 | 1ADS | 74 |
| ATOM | 6 | N | SER | 2 | 14.081 | 42.818 | 84.000 | 1.00 | 42.59 | 1ADS | 75 |
| ATOM | 7 | CA | SER | 2 | 14.302 | 41.696 | 83.095 | 1.00 | 39.18 | 1ADS | 76 |
| ATOM | 8 | C | SER | 2 | 13.293 | 41.837 | 81.942 | 1.00 | 33.16 | 1ADS | 77 |
| ATOM | 9 | O | SER | 2 | 13.093 | 40.795 | 81.346 | 1.00 | 32.04 | 1ADS | 78 |
| ATOM | 10 | CB | SER | 2 | 14.082 | 40.280 | 83.820 | 1.00 | 42.04 | 1ADS | 79 |
| ATOM | 11 | OG | SER | 2 | 14.246 | 39.016 | 83.073 | 1.00 | 43.89 | 1ADS | 80 |
| ATOM | 12 | N | ARG | 3 | 12.600 | 42.930 | 81.524 | 1.00 | 26.98 | 1ADS | 81 |
| ATOM | 13 | CA | ARG | 3 | 11.602 | 42.867 | 80.474 | 1.00 | 21.66 | 1ADS | 82 |
| ATOM | 14 | C | ARG | 3 | 11.690 | 44.098 | 79.628 | 1.00 | 19.49 | 1ADS | 83 |
| ATOM | 15 | O | ARG | 3 | 12.001 | 45.184 | 80.116 | 1.00 | 18.09 | 1ADS | 84 |
| ATOM | 16 | CB | ARG | 3 | 10.185 | 42.775 | 81.017 | 1.00 | 24.23 | 1ADS | 85 |
| ATOM | 17 | CG | ARG | 3 | 9.917 | 41.399 | 81.655 | 1.00 | 29.74 | 1ADS | 86 |
| ATOM | 18 | CD | ARG | 3 | 8.485 | 40.971 | 81.973 | 1.00 | 35.55 | 1ADS | 87 |
| ATOM | 19 | NE | ARG | 3 | 7.911 | 41.674 | 83.116 | 1.00 | 41.64 | 1ADS | 88 |
| ATOM | 20 | CZ | ARG | 3 | 6.581 | 41.695 | 83.396 | 1.00 | 44.91 | 1ADS | 89 |
| ATOM | 21 | NH1 | ARG | 3 | 5.690 | 41.075 | 82.575 | 1.00 | 46.93 | 1ADS | 90 |
| ATOM | 22 | NH2 | ARG | 3 | 6.150 | 42.355 | 84.498 | 1.00 | 46.03 | 1ADS | 91 |

**Table 1.**

Table 1 shows a portion of the spherical coordinate system structure file (top half) of a protein, and the PDB coordinate (in Cartesian system) file (lower half) from which it was derived. The first five columns are identical, but the sixth, seventh and eighth columns differ; in the PDB file, these are the x-, y- and z-coordintes, respectively, and in the spherical coordinate system file, these are the rho ($\rho$), phi ($\varphi$) and theta ($\theta$) coordinates, where $\varphi$ and are angles in degress or radians (in this case, degrees). The ninth and tenth columns in the PDB file are the positional occupancies and B-factors, respectively.

**Table 2.**

| Protein | Abbrev.[1] | PDB ID | # ligands | Description |
|---|---|---|---|---|
| A | Ae | 1TDE | 1 | Thioredoxin Reductase from *E. coli* |
| B | Au | 1RPA | 3 | Rat Acid Phosphatase |
| C | Bc | 1XNB | 1 | Xylanases from *B. Circulans* and *T. Harzianum* |
| D | Bm | 1HNE | 5 | Human Neutrophil Elastase |
| E | Cg | 1PDA | 2 | Porphobilinogen Deaminase from *E. coli* |
| F | Co | 4BCL[2] | 7 | Fenna–Matthews–Olson protein from *P. aestuarii* |

[1] in reference paper, Reyes, V.M., 2015x
[2] same as protein 3EOJ (FMO protein from *P. aestuarii)*

Table 2 shows the identities of the six test proteins, A – F, we used to show how the Fortran programs presented here work and are implemented. The second column shows the abbreviations used in the main paper (Reyes, V.M., 2015a) to refer to these proteins. These are six of the 67 test proteins from the dataset of Laskowski et al., 1996. Note that they are all bound (liganded) forms of the protein in question, and that the number of bound ligands vary. Bound ligand in protein A is FAD; ligands TAR and two NAG molecules in protein B; SO4 in protein C; ALM, MSU, two ALA and one PRO molecules in protein D; DPM and ACY in protein E; and seven molecules of BCL in protein F. Please go to the PDB website (www.rcsb.org) for the molecular identities of these ligands.

**All- Atom Representation, Fine Binning**

| Protein | OUTER LAYER | | INNER CORE | |
|---|---|---|---|---|
| | %Hphi | %Hpho | %Hphi | %Hpho |
| A | 70% | 29% | 18% | 57% |
| B | 68% | 31% | 19% | 57% |
| C | 78% | 21% | 14% | 61% |
| D | 55% | 44% | 8% | 68% |
| E | 61% | 38% | 18% | 58% |
| F | 68% | 32% | 28% | 49% |

**Table 3 A.**

**All- Atom Representation, Coarse Binning**

| Protein | OUTER LAYER | | INNER CORE | |
|---|---|---|---|---|
| | %Hphi | %Hpho | %Hphi | %Hpho |
| A | 78% | 26% | 19% | 54% |
| B | 72% | 30% | 20% | 53% |
| C | 80% | 22% | 14% | 60% |
| D | 54% | 49% | 8% | 67% |
| E | 68% | 36% | 17% | 59% |
| F | 70% | 33% | 27% | 48% |

**Table 3 B.**

**Double-Centroid Reduced Representation, Coarse Binning**

| Protein | OUTER LAYER | | INNER CORE | |
|---|---|---|---|---|
| | %Hphi | %Hpho | %Hphi | %Hpho |
| A | 80% | 20% | 39% | 73% |
| B | 72% | 28% | 40% | 72% |
| C | 85% | 17% | 46% | 68% |
| D | 53% | 47% | 34% | 79% |
| E | 63% | 38% | 39% | 74% |
| F | 67% | 34% | 47% | 68% |

**Table 3 C.**

Table 3, Panels A, B and C are related to the plots shown in Figure 1, Panels A, B and C, resoectively. Panel A shows the results when fine binning is used on the proteins in all-atom representation; Panel B, shows the results when coarse binning is used on the proteins in all-atom representation; and Panel C shows the results when coarse binning is used on the proteins in double-centroid reduced representation. Note that in every case, there is a significantly higher % of hydrophilic residues than hydrophobic ones in the OL, while exactly the reverse is true for the IC.

| PDB ID | Protein Description | Number of residues predicted to be in and around active site(s) | |
|---|---|---|---|
| | | coarse binning | fine binning |
| 1TDE | Thioredoxin reductase | 44 | 33 |
| 1RPA | Prostatic acid phosphatase | 53 | 36 |
| 1XNB | Xylanase | 5 | 2 |
| 1HNE | Human neutrophil elastase | 8 | 4 |
| 1PDA | Porphobilinogen deaminase | 41 | 21 |
| 4BCL | Bacteriochlorophyll-A protein | 41 | 28 |

**Table 4.**

Table 4 shows the results for the six test proteins after the procedure was applied to them to analyze their surface topographies as to the presence of invaginations, which would indicate a potential active site or ligand binding site. The predictions largely match the visual analysis of the surfaces of the proteins as found by Laskowski et al. 1996. The results for all 67 test proteins in the original dataset by Laskowski et al. 1996 are found in our main paper describing the methodology (Reyes, V.M., 2011).